\documentclass[a4paper,twocolumn,aps]{revtex4}
\usepackage[T1]{fontenc}
\usepackage{graphics}

\makeatletter

\providecommand{\LyX}{L\kern-.1667em\lower.25em\hbox{Y}\kern-.125emX\@}
\let\SF@@footnote\footnote
\def\footnote{\ifx\protect\@typeset@protect
    \expandafter\SF@@footnote
  \else
    \expandafter\SF@gobble@opt
  \fi
}
\expandafter\def\csname SF@gobble@opt \endcsname{\@ifnextchar[%]
  \SF@gobble@twobracket
  \@gobble
}
\edef\SF@gobble@opt{\noexpand\protect
  \expandafter\noexpand\csname SF@gobble@opt \endcsname}
\def\SF@gobble@twobracket[#1]#2{}

\usepackage[T1]{fontenc}
\usepackage[latin1]{inputenc}
\usepackage{color}
\usepackage{graphics}

\makeatletter

\usepackage{latexsym}

\makeatother
\makeatother

\begin{document}

\preprint{This paper is printed with \textbf{preprint} option}

\title{Effects of a dipolar field in the spin dynamics of a Fermi liquid}

\author{P. L. Krotkov}

\email{krotkov@itp.ac.ru}

\address{L. D. Landau Institute for Theoretical Physics, Russian Academy of Sciences,
2 Kosygina st., 117334 Moscow, Russia}

\author{V. P. Mineev}

\email{mineev@drfmc.ceng.cea.fr}

\address{D\'{e}partament de Recherche Fondamentale sur la Mati\`{e}re Condens\'{e}e,
Commisariat \`{a} l'\'{E}nergie Atomique, 17 rue des Martyrs, 38054 Grenoble
Cedex 9, France}

\author{G. A. Vermeulen}

\email{gvermeul@labs.polycnrs-gre.fr}

\homepage{http://www-crtbt.polycnrs-gre.fr/he3pol}

\address{Centre des Recherches sur les Tr\`{e}s Basses Temp\'{e}ratures, laboratoire
associ\'{e} \`{a} l'Universit\'{e} Joseph Fourier, CNRS, BP 166, Grenoble
C\'{e}dex 9, France}

\begin{abstract}
We study spin dynamics of a normal Fermi liquid taking into account the demagnetizing
field produced by the spin system itself. Linear solutions of the spin dynamics
equations in the form of standing spin waves in a finite volume of liquid are
found. At almost all known experimental conditions the influence of demagnetizing
field can be satisfactorily described by the first order of perturbation theory.
We carried out perturbational calculations for two geometries of experimental
cell --- spherical and finite-cylindrical. We performed also exact numerical
simulations of the spin wave spectra in a spherical cell at an arbitrary strength
of the demagnetizing field.

The obtained results are applied in particular to conditions of recent experiment
(G.Vermeulen and A.Roni, Phys. Rev. Lett. \textbf{86}, 248 (2001)) related to
the problem of zero temperature transverse relaxation in a polarized Fermi liquid.
We found that not taking into account demagnetizing field leads to negligible
errors in the measured relaxation time, thus supporting the conclusion of the
absence of zero temperature spin wave damping.

\end{abstract}

\date{January, 20, 2001}

\maketitle
\newcommand{\f}[1]{\mbox {\boldmath\(#1\)}}

\newcommand{\ds}[1]{\displaystyle #1}

\newcommand{\ts}[1]{\textstyle #1}

\newcommand{\rev}{\rangle \! \! \rangle \! \! \rangle \! \! \rangle }

\newcommand{\lev}{\langle \! \! \langle \! \! \langle \! \! \langle }

\section{Introduction}

Spin dynamics of a strongly spin-polarized normal Fermi liquid still captures
appreciable theoretical and experimental interest. Among the main questions
here is whether the transverse (i.e. in a direction perpendicular to the external
field) magnetization excitations are damped at zero temperature.

Polarizing a Fermi liquid creates a gap \( \sim \hbar \gamma M/\chi _{n} \)
between the two Fermi energies for spins up and down. Here \( \gamma  \) is
the gyromagnetic ratio of \( ^{3} \)He nuclei, \( \chi _{n} \) is the susceptibility
of the liquid. Meyerovich has pointed out \cite{Meyerovich} that the existence
of the gap leads to a non-conventional temperature dependence of the transverse
relaxation time, \( \propto (T^{2}+T^{2}_{\mathrm{a}})^{-1} \), where \( T_{\mathrm{a}} \)
is of the order of the gap, and therefore to a damping of the transverse excitations
even at zero temperature. This idea has been pursued in several theoretical
papers \cite{JeonMullin}, \cite{MeyerovichMusaelian}, \cite{Golosov} and
was contested recently by Fomin \cite{Fomin-solo}, who argued that the conclusion
of existence of zero-temperature attenuation is drawn from the wrong premises
about the ground state of a polarized Fermi-liquid, viz., from treating the
quasiparticles between two Fermi levels as excitations. Whereas as long as the
polarization of the liquid is conserved, these particles should be considered
as an inalienable part of the ground state.

Curiously, a similar discussion arose in mesoscopic physics where a poorly argued
concept of finite dephasing time at \( T=0 \) has been developed \cite{golubev98}
and contested \cite{aleiner99}, \cite{cohen99}.

The results of the measurements of the spin diffusion coefficients by spin echo
experiments in pure \( ^{3} \)He \cite{Wei}, \cite{candela  OB} and in solutions
of \( ^{3} \)He in \( ^{4} \)He \cite{Ager}, \cite{OB} with a spin polarization
of a few percent revealed a finite value of \( T_{\mathrm{a}} \) in a qualitative
agreement with the zero-attenuation concept. But the observed \( T_{\mathrm{a}} \)
were several times greater than theoretical estimations in \cite{JeonMullin},
\cite{MeyerovichMusaelian}, \cite{Golosov}. On the contrary, the recent measurements
of linear spin wave damping in dilute \( ^{3} \)He at even higher polarizations
\cite{vermeulen} are in agreement with Fomin's theory (\( T_{\mathrm{a}}=0 \)),
although the upper limit for a finite \( T_{\mathrm{a}} \) set by this experiment
(due to the error bars) does not allow to rule out completely the existing theory
of zero temperature spin wave damping. 

A coherent theory of strongly polarized Fermi liquids based on a properly defined
ground state is lacking. On the other hand a proper interpretation of the experimental
data for a strongly polarized liquid is itself non-trivial.

The point is that the magnetic field acting on the spins of a liquid is conventionally
supposed to equal the external field \( \mathbf{H}^{e} \). In reality the field
inside a specimen is well-known to differ from \( \mathbf{H}^{e} \) due to
the shape-dependent demagnetizing field proportional to the magnetization. An
oscillating magnetization thus acts back on itself via the demagnetizing (or
dipolar) field. This phenomenon manifests itself as magnetostatic waves in ferrimagnets
theoretically described by Kittel \cite{kittel} and Walker \cite{walker}.
Walker originally showed that if the magnetization \( \mathbf{M} \) is supposed
to obey the Larmor precession around the internal magnetic field \textbf{\( \mathbf{H}^{i} \)
\begin{equation}
\label{precession}
(\partial _{t}+\gamma \mathbf{H}^{i}\times )\mathbf{M}=0,
\end{equation}
}then because \( \mathbf{M} \) and \( \mathbf{H}^{i} \) are related through
Maxwell equations they can self-consistently oscillate only with certain frequencies
localized in the range \( \sim 2\pi \gamma M \) near the Larmor frequency.

In an interacting Fermi liquid (\ref{precession}) should be replaced with the
Leggett system of equations. In the linear approximation, solutions of this
system are standing spin wave modes with the widths proportional to the transverse
relaxation time. So study of the behavior of the widths of the modes with temperature
is one of the possible ways to detect the zero-temperature transverse attenuation.

For \emph{weakly-polarized} liquids the effects of the demagnetizing field can
be discarded. But it is preferable to have \emph{strong polarizations} in order
to increase the predicted temperatures of the attenuation onset. So the effect
of polarization has to be taken into consideration for the proper interpretation
of the spectra. On the other hand at sufficiently strong polarizations one can
expect considerable changes in the Leggett description of the Fermi liquid spin
dynamics. In particular, due to the presence of two Fermi surfaces a double
set of the Fermi-liquid parameters must enter the theory. We will use Leggett
equations assuming that they are still valid when the share of polarized nuclei
of the liquid does not surpass \( \sim  \)10 \%.

To include dipolar field we have chosen to write out the dipolar part of the
internal field \( \mathbf{H}^{i} \) explicitly as an integral of the magnetization,
this integral being a general solution of Maxwell differential equations with
Maxwell boundary conditions. Thus we work with a closed integro-differential
equation directly on the magnetization.

A short review of other possible methods is done in the Discussions section.

We start directly from the generalized Leggett equations and study the effects
of the demagnetizing field coherently. We specialize to the case of linear spin
waves, setting spin echo experiments aside.

A full-blown study including numerical calculations is done only for a spherical
shape. For a finite cylinder as well as for a sphere we also calculated the
changes to the spectra by the demagnetizing field using perturbation theory.

Our results show that the demagnetizing field introduces small corrections (about
4\%) to the value of low temperature transversal relaxation rate experimentally
determined by G.Vermeulen and A.Roni \cite{vermeulen}. Hence, the main conclusion
of Ref. \cite{vermeulen} about the absence of zero-temperature spin wave damping
is supported.

The paper is organized as follows. Sections II and III form the basis needed
to comprehend the authors' point of view. In Sec. II we show how to include
the dipolar field in the standard Leggett equations and to how to linearize
the result to obtain an equation for spin waves subject to both exchange and
demagnetizing fields. 

In Sec. III as the simplest application of the theory developed we find corrections
to the spin wave spectra in a finite cylinder in the first order of perturbation
theory. At the end of the Sec. III we find dipolar limitations on the correct
determination of the transverse relaxation time from the conventional interpretation
of the spectra.

The next Section IV contains similar first order perturbational estimation of
the dipolar-field corrections to the spin waves spectra in a spherical container.

The results for a sphere are compared to numerical simulations carried out in
Section V, where we also calculate spin wave spectra in the regimes of intermediate
and strong demagnetizing fields.

In the last Section VI we discuss conclusions.

\section{Statement of the problem}

\subsection{Basic equations}

Spin dynamics of a Fermi liquid is described by the Leggett coupled system of
two partial differential equations on the local magnetization \( \mathbf{M}(\mathbf{r},t) \)
and its current \( \mathbf{J}_{i}(\mathbf{r},t) \) \cite{Leggett}

\begin{eqnarray}
\left( \partial _{t}+\gamma \mathbf{B}\times \right)  & \mathbf{M} & +\partial _{i}\mathbf{J}_{i}=0,\label{laggett-1} \\
\left( \partial _{t}+\gamma \mathbf{B}\times \right)  & \mathbf{J}_{i} & +\frac{w^{2}}{3}\partial _{i}(\mathbf{M}-\mathbf{M}_{0})\label{laggett-2} \\
 & + & \kappa \frac{\gamma }{\chi _{n}}\mathbf{M}\times \mathbf{J}_{i}=-\frac{\mathbf{J}_{i}}{\tau _{1}}.\nonumber 
\end{eqnarray}

Here \( \mathbf{B} \) is the flux density inside the sample \cite{20}, \( w^{2} \)
the renormalized Fermi velocity \( w^{2}=v_{F}^{2}(1+F^{a}_{0})(1+F^{a}_{1}/3) \)
and \( \tau _{1} \) the renormalized relaxation time \( \tau _{1}=\tau /(1+F^{a}_{1}/3) \).
\( F^{a}_{0} \) and \( F^{a}_{1} \) are the coefficients of expansion of the
antisymmetric part of the Fermi-liquid interaction in the spherical harmonics.
The equilibrium magnetization is 
\begin{equation}
\label{equilibrium-mag}
\mathbf{M}_{0}=\chi _{n}\mathbf{H}^{i},
\end{equation}
 where \( \mathbf{H}^{i}=\mathbf{B}-4\pi \mathbf{M} \) is the internal field.

Spin dynamics equations of a Fermi liquid reduce to the form (\ref{laggett-1}),
(\ref{laggett-2}) in either of the regimes --- collisionless \( C\gg 1 \)
or hydrodynamic \( C\ll 1 \), where the regime parameter
\begin{equation}
\label{regimeC}
C=\kappa (\gamma H^{i})(M/M_{0})\tau _{1}.
\end{equation}
 The factor \( M/M_{0} \) accounts for the possibility for the polarization
\( M \) be higher than the equilibrium value \( M_{0} \) --- in experiments
\cite{vermeulen}, \cite{Roni} \( M/M_{0} \) varied from \( 1 \) to \( 5 \).

The condition of applicability of the Leggett equations \cite{Leggett} is that
the characteristic scale of spatial inhomogeneity \( \xi  \) be greater than
the quasiparticle mean free path \( v_{F}\tau  \) \emph{or} the magnetic length
\( v_{F}\tau /C \) which one is the shorter, 
\begin{equation}
\xi \gg \min \left\{ v_{F}\tau ,\, v_{F}\tau /C\right\} .
\end{equation}
 In the collisionless regime (\( C\gg 1 \)) the magnetic length \( v_{F}\tau /C \)
is the shorter and thus should be smaller than \( \xi  \). While in the hydrodynamic
regime (\( C\ll 1 \)) the spatial scale \( \xi  \) should exceed the mean
free path \( v_{F}\tau  \).

Eq.~(\ref{laggett-2}) contains the torque due to the \emph{local} molecular
field \( \kappa \mathbf{M}(\mathbf{r},t)/\chi _{n} \) acting on current \( \mathbf{J}_{i}(\mathbf{r},t) \).
The combination of the Fermi-liquid constants \( F^{a}_{0} \) and \( F^{a}_{1} \)
\begin{equation}
\kappa =\frac{\frac{1}{3}F^{a}_{1}-F^{a}_{0}}{1+F^{a}_{0}}
\end{equation}
 measures the strength of the exchange interaction. It vanishes when turning
the exchange off.

Leggett originally \cite{Leggett} considered the case of a weakly-polarized
sample, wherein \( \mathbf{B}=\mathbf{H}^{e} \) --- the external magnetic field
at infinity. Generally, the relation between \( \mathbf{B} \) and \( \mathbf{H}^{e} \)
is to be determined from the conventional boundary value problem of solving
the magnetostatic equations in a non-conducting medium
\begin{eqnarray}
\f \partial \times \mathbf{H} & = & 0,\qquad \f \partial \mathbf{B}=0\label{ms} 
\end{eqnarray}
 with Maxwell boundary conditions of the continuity of \( B_{n} \) and \( \mathbf{H}_{t} \)
at the boundary of the sample and of \( \mathbf{H}\rightarrow \mathbf{H}^{e} \)
at infinity. In (\ref{ms})
\begin{equation}
\label{B-dip-2}
\mathbf{H}=\mathbf{B}-4\pi \mathbf{M}.
\end{equation}

A general formal solution of (\ref{ms}), (\ref{B-dip-2}) satisfying the appropriate
boundary conditions is \cite{Jackson}
\begin{eqnarray}
\mathbf{H} & = & \mathbf{H}^{e}+\mathbf{H}_{\mathrm{dip}},\label{B-dip-1} 
\end{eqnarray}
 where
\begin{equation}
\label{B-dip}
\mathbf{H}_{\mathrm{dip}}(\mathbf{r})=\f \partial \left( \f \partial \int \frac{\mathbf{M}(\mathbf{r}')}{|\mathbf{r}-\mathbf{r}'|}d^{3}\mathbf{r}'\right) 
\end{equation}
 is called the dipolar field. It is straightforward to verify that Eqs.~(\ref{B-dip-1})
and (\ref{B-dip-2}) are indeed the solution in the whole space with the help
of
\begin{equation}
\label{green-inf}
\partial ^{2}|\mathbf{r}-\mathbf{r}'|^{-1}=-4\pi \delta (\mathbf{r}-\mathbf{r}').
\end{equation}

\( \mathbf{M} \) in its turn has to be found from the Leggett equations (\ref{laggett-1}),
(\ref{laggett-2}). Therefore, the closed system of integro-differential Eqs.~(\ref{laggett-1}),
(\ref{laggett-2}), (\ref{B-dip-2})--(\ref{B-dip}) completely describes normal
Fermi liquid electrodynamics with the effects of both inhomogeneity and the
demagnetizing field taken into account.

Inside the sample the field \( \mathbf{H} \) from (\ref{B-dip-1}) is called
\( \mathbf{H}^{i} \) --- the internal magnetic field. The difference between
external field at infinity \( \mathbf{H}^{e} \) and \( \mathbf{H}^{i} \) is
usually denoted as 
\begin{equation}
\label{Hi-He}
\mathbf{H}^{i}-\mathbf{H}^{e}=-4\pi \underline{\widehat{n}}[\mathbf{M}],
\end{equation}
 where the (tensor) operator \( \underline{\widehat{n}} \), acting by the rule
\begin{equation}
\label{demag-tensor}
\underline{\widehat{n}}[\mathbf{M}]=-\frac{1}{4\pi }\f \partial \left( \f \partial \int _{V}\frac{\mathbf{M}(\mathbf{r}')}{|\mathbf{r}-\mathbf{r}'|}d^{3}\mathbf{r}'\right) 
\end{equation}
 is called demagnetizing operator. Let us agree to denote an operator with a
hat over a letter and a tensor with an underlined letter. The demagnetizing
tensor operator \( \underline{\widehat{n}} \) is specific to the shape of the
sample, over volume of which the integration in (\ref{demag-tensor}) is taken,
and generally is coordinate-dependent. It only reduces to the tensor of constant
demagnetizing coefficients \( \underline{n} \) when acting on spatially homogeneous
distributions in ellipsoidal samples (including limiting cases of a slab and
an infinite cylinder).

\subsection{Static magnetization distribution}

In a linear spin wave the magnetization rotates with a small amplitude \( \mathbf{m} \)
around its (large) stable value \( \mathbf{M} \) in a static external magnetic
field \( \mathbf{H}^{e}=H^{e}\hat{z} \).

Consider a general static (\( \partial _{t}\mathbf{M}=0 \), \( \mathbf{J}_{i}=0 \))
case. In order for \( \mathbf{J}_{i}=0 \) there should be 
\begin{equation}
\label{dM-M0}
\partial _{i}(\mathbf{M}-\mathbf{M}_{0})=0
\end{equation}
 and then for \( \partial _{t}\mathbf{M}=0 \) the magnetization should be locally
directed along \( \mathbf{H}^{i} \)
\begin{equation}
\label{MonHi}
\mathbf{M}(\mathbf{r})=\chi _{n}A(\mathbf{r})\mathbf{H}^{i}(\mathbf{r}).
\end{equation}
 The function \( A(\mathbf{r}) \) must be such that \( \mathbf{M}(\mathbf{r}) \)
satisfies (\ref{dM-M0}). This imposes restrictions on the spatial dependence
of \( A(\mathbf{r}) \), but leaves \( A(0) \) arbitrary.

So there exists continuum of static non-equilibrium magnetization distributions
numbered by \( A\equiv A(0) \). It is this \( A \) which is represented as
\( M/M_{0} \) in Table I.

In equilibrium (\ref{equilibrium-mag}) \( A(\mathbf{r})\equiv 1 \).

To find a form of a non-equilibrium static magnetization distribution, we shall
use the smallness of \( \chi _{n} \). For pure \( ^{3} \)He, the magnetic
susceptibility is 
\begin{equation}
\chi _{n}=\hbar ^{2}\gamma ^{2}N_{0}/2(1+F^{a}_{0})\sim 10^{-7},
\end{equation}
 where \( N_{0}=m^{*}k_{F}/2\pi ^{2}\hbar ^{2} \) is the density of states
on the Fermi surface. For \( ^{3} \)He-\( ^{4} \)He mixtures \( \chi _{n} \)
is less, proportional to \( k_{F}\propto \sqrt[3]{x} \), where \( x \) is
the concentration of \( ^{3} \)He atoms in the mixture.

Substituting (\ref{MonHi}) into (\ref{Hi-He}) yields 
\begin{equation}
\label{Hi}
\mathbf{H}^{i}(\mathbf{r})=\mathbf{H}^{e}(\mathbf{r})-4\pi \chi _{n}\underline{\widehat{n}}\left[ A(\mathbf{r})\mathbf{H}^{e}(\mathbf{r})\right] +O(\chi ^{2}_{n}).
\end{equation}

For the reasons that will become clear below the external magnetic field is
taken almost constant, with a small gradient along its direction
\begin{equation}
\label{He}
\mathbf{H}^{e}(\mathbf{r})=\mathbf{H}^{e}(1+z\nabla \omega _{L}/\omega _{L}).
\end{equation}
 The presence of the field gradients in the perpendicular directions, necessary
for the fulfillment of the condition \( \f \partial \mathbf{H}^{e}=0 \), is
inessential for the following discourse.

From (\ref{dM-M0}) it then follows that \( \nabla A=(1-A)\nabla \omega _{L}/\omega _{L} \),
i.e. the spatial inhomogeneity of \( A(\mathbf{r})=A+z\nabla A \) has the same
smallness.

Leaving in (\ref{Hi}) only the first order terms in either \( \chi _{n} \)
or \( z\nabla \omega _{L}/\omega _{L} \), we have 
\begin{eqnarray}
\gamma \mathbf{H}^{i} & = & (\omega _{L}+z\nabla \omega _{L})\hat{z}-\omega _{M}\underline{\widehat{n}}[\hat{z}].\label{HiM} 
\end{eqnarray}
 Here 
\begin{equation}
\omega _{L}=\gamma H^{e}
\end{equation}
 is the Larmor frequency and 
\begin{equation}
\omega _{M}=4\pi \gamma M=4\pi \chi _{n}A\omega _{L}
\end{equation}
 is the frequency corresponding to magnetization. \( \underline{\widehat{n}}[\hat{z}] \)
is generally a coordinate dependent vector. For ellipsoidal samples it reduces
to \( \underline{n}\hat{z} \). If one of the principal axes of the ellipsoid
(which are also the principal axes of the demagnetizing coefficients tensor
\( \underline{n} \)) coincides with \( \hat{z} \), we have \( \underline{n}\hat{z}=\hat{z}n^{(z)} \),
where \( n^{(z)} \) is the \( z \)-th demagnetizing coefficient. E.g., for
a sphere \( n^{(z)}=\frac{1}{3} \), for a plane-parallel slab with the edges
perpendicular to \( \widehat{z} \) the coefficient \( n^{(z)}=1 \), and for
an infinite circular cylinder with the generatrix parallel to \( \widehat{z} \)
the coefficient \( n^{(z)}=0 \).

The small ratio \( 2R\nabla \omega _{L}/\omega _{L} \), where \( 2R \) is
the sample size, together with \( 4\pi \chi _{n}A(0)\sim 10^{-6} \) are the
two small parameters in the problem. Conventionally the dipolar field is not
taken into account and the second parameter is considered negligibly small.
This is no longer justified for recent experimental conditions as is seen from
Table I, where the ratio of the second parameter to the first 
\begin{equation}
\omega _{M}/2R\nabla \omega _{L}
\end{equation}
 is represented for various experimental conditions.

\subsection{Linearized equations of motion}

To obtain the linearized form of the equations of motion (\ref{laggett-1}),
(\ref{laggett-2}) we expand all the macroscopic quantities near their stationary
values: 
\begin{eqnarray}
\mathbf{H}^{e}=H^{e}\widehat{z}+\mathbf{h}^{e}, & \quad  & \mathbf{H}^{i}=\mathbf{H}^{i}+\mathbf{h}^{i},\nonumber \\
\mathbf{M}=\mathbf{M}+\mathbf{m}, & \quad  & \mathbf{J}_{i}=\mathbf{j}_{i},\label{linear} 
\end{eqnarray}
 where the static value of \( \mathbf{H}^{i} \) is that from (\ref{HiM}).
The radio frequency field \( \mathbf{h}^{e} \) plays the role of a driving
force for the spin system response \( \mathbf{m} \). We denote \( \mathbf{h}^{i}=\mathbf{h}^{e}-4\pi \underline{\widehat{n}}[\mathbf{m}] \).

The static \( \mathbf{M} \) in (\ref{linear}) is the result of the substitution
of (\ref{HiM}) into (\ref{MonHi}) 
\begin{equation}
\label{Mmoreaccu}
\gamma \mathbf{M}/\chi _{n}=\omega _{L}(A+z\nabla \omega _{L})\hat{z}-\omega _{M}AH^{e}\underline{\widehat{n}}[\hat{z}].
\end{equation}
 However, we can retain only the greatest term in \( \mathbf{M} \)
\begin{equation}
\label{MHi}
\mathbf{M}=\chi _{n}A\mathbf{H}^{e}
\end{equation}
 when linearizing Eqs.~(\ref{laggett-1}), (\ref{laggett-2}). In (\ref{laggett-1})
this is simply due to the fact that the terms of \( \mathbf{M} \) to be omitted
are of the order \( O(\chi _{n}z\nabla \omega _{L}/\omega _{L}) \) and \( O(\chi ^{2}_{n}) \).
In (\ref{laggett-2}) \( \mathbf{M} \) is divided by \( \chi _{n} \) and the
justification is lengthier. We will suppose (\ref{MHi}) and discuss why only
the main term in \( \mathbf{M} \) should be left after the derivation below.

In practice one usually is interested in movements quasistationary in the Larmor
frame. To a first approximation one supposes that \( \mathbf{j}_{i}(\mathbf{r},t) \)
is stationary, i.e. precesses with the frequency \( \gamma \mathbf{B} \): \textbf{\( (\partial _{t}+\gamma \mathbf{B}\times )\mathbf{j}_{i}=0 \)}.
Then resolving Eq.~(\ref{laggett-2}) with respect to \( \mathbf{j}_{i} \)
with \( \mathbf{M} \) from (\ref{MHi}) gives
\begin{equation}
\label{j-i}
\mathbf{j}_{i}=-\frac{D_{0}}{1+C^{2}}\left[ \partial _{i}\mathbf{m}-C\widehat{z}\times \partial _{i}\mathbf{m}+C^{2}\widehat{z}(\widehat{z}\partial _{i}\mathbf{m})\right] ,
\end{equation}
 where the diffusion coefficient \( D_{0}=\frac{1}{3}w^{2}\tau _{1}=\frac{1}{3}v_{F}^{2}\tau (1+F^{a}_{0}) \)
and
\begin{equation}
C=\kappa \tau _{1}\gamma M/\chi _{n}=\kappa \tau _{1}A\omega _{L}
\end{equation}
 is another expression for the regime parameter (\ref{regimeC}).

One then plugs the divergence of (\ref{j-i}) into (\ref{laggett-1}). The divergence
of \( \mathbf{j}_{i} \) has the order 
\begin{equation}
\partial _{i}\mathbf{j}_{i}\sim D_{0}\partial ^{2}\mathbf{m}\sim (D_{0}/\xi ^{2})\mathbf{m},
\end{equation}
 where \( \xi \sim \sqrt[3]{D_{0}/\nabla \omega _{L}} \) is the characteristic
scale (\ref{lambda}). So \( \partial _{i}\mathbf{j}_{i}\sim (\xi \nabla \omega _{L})\mathbf{m} \)
\emph{has already the smallness} \( z\nabla \omega _{L}/\omega _{L} \). If
we had accounted for the higher order terms in \( \mathbf{M} \) than (\ref{MHi})
when calculating the current, these terms would have entered \( \mathbf{j}_{i} \)
through the regime parameter \( C \), and after multiplication with \( D_{0}\partial ^{2}\mathbf{m} \)
would have produced terms of the order \( O(\chi _{n}z\nabla \omega _{L}/\omega _{L})\mathbf{m} \)
and \( O(\chi ^{2}_{n})\mathbf{m} \). That is why we were allowed to substitute
simply (\ref{MHi}) in (\ref{laggett-2}).

In the linear approximation \( \mathbf{m}(\mathbf{r}) \) is in each point perpendicular
to the static \( \mathbf{M}(\mathbf{r}) \) if the absolute value of the magnetization
is conserved. Exp. (\ref{Mmoreaccu}) shows that apart from the major \( \hat{x}- \)
and \( \hat{y}- \) components \( \mathbf{m} \) also has a minor \( m_{z}\sim \chi _{n}m_{x} \).
This component also may be seen to give higher order terms and is therefore
negligible.

We will thus consider \( \mathbf{m}\bot \widehat{z} \). The last term on the
right-hand side of (\ref{j-i}) then vanishes, and substituting (\ref{j-i})
into (\ref{laggett-1}) one gets 
\begin{equation}
\label{torque-term}
\left( \partial _{t}-\ts {\frac{D}{C}}\partial ^{2}\right) \mathbf{m}+\widehat{z}\times \left( \gamma H_{z}^{i}\mathbf{m}+D\partial ^{2}\mathbf{m}-\gamma M\mathbf{h}^{i}\right) =0.
\end{equation}
 Here we left only the \( \hat{z}- \)component of \( \mathbf{H}^{i} \) because
\( \mathbf{H}_{\perp }^{i} \), which multiplies vectorially only by \( m_{z}\hat{z} \),
produces terms \( O(\chi ^{2}_{n}) \).

In (\ref{torque-term}) we introduced the effective spin diffusion coefficient
\begin{equation}
D=D_{0}C/(1+C^{2}).
\end{equation}
 In the strong (\( C\gg 1 \)) collisionless regime \( D\approx w^{2}/3\kappa \omega _{L}A \)
is temperature independent.

Suppose \( \mathbf{m} \), \( \mathbf{h} \) depend on time as monochromatic
waves \( \mathbf{m} \), \( \mathbf{h}\propto e^{-i\omega t} \). Then written
out in components the equation of motion for \( \mathbf{m} \) becomes 
\begin{equation}
-i\widehat{\omega }\left( m_{x}\atop m_{y}\right) +\widehat{\omega }_{L}\left( -m_{y}\atop m_{x}\right) =\frac{\omega _{M}}{4\pi }\left( -h^{i}_{y}\atop h^{i}_{x}\right) ,
\end{equation}
 where we designated by \( \widehat{\omega }_{L} \) and \( \widehat{\omega } \)
respectively the operators 
\begin{equation}
\widehat{\omega }_{L}=\gamma H_{z}^{i}+D\partial ^{2},\quad \widehat{\omega }=\omega -i\frac{D}{C}\partial ^{2}.
\end{equation}

Multiplying by 
\[
\left( \begin{array}{cc}
0 & 1\\
-1 & 0
\end{array}\right) \]
 results in the second order inhomogeneous partial differential equation \( \underline{\widehat{\chi }}^{-1}_{\omega }\mathbf{m}=\mathbf{h}^{i}_{\bot } \),
or
\begin{equation}
\label{motion}
\underline{\widehat{\chi }}^{-1}_{\omega }\mathbf{m}+4\pi \underline{\widehat{n}}[\mathbf{m}]=\mathbf{h}^{e}_{\bot }.
\end{equation}
 Here \( \underline{\widehat{\chi }}^{-1}_{\omega } \) is the inverse susceptibility
tensor 
\begin{equation}
\underline{\widehat{\chi }}^{-1}_{\omega }=\frac{4\pi }{\omega _{M}}\left( \begin{array}{cc}
\widehat{\omega }_{L} & -i\widehat{\omega }\\
i\widehat{\omega } & \widehat{\omega }_{L}
\end{array}\right) .
\end{equation}

To close the boundary value problem one must impose some appropriate boundary
conditions on \textbf{\( \mathbf{m} \)}. Supposed that the container is made
from a non-magnetic material, there is no magnetization current into the walls,
and we get boundary condition in the form 
\begin{equation}
\label{bc}
\left. \hat{n}_{i}\partial _{i}\mathbf{m}\right| _{\partial }=0.
\end{equation}
 Here \( \hat{n}_{i} \) is a unit normal to the wall.

In the normal variables \( m_{\pm }=m_{x}\pm im_{y} \) Eq.~(\ref{motion})
has equivalent form 
\begin{eqnarray}
(\widehat{\omega }_{L}\mp \widehat{\omega })m_{\pm } & = & \frac{\omega _{M}}{4\pi }\left( h_{\pm }^{e}+\partial _{\pm }\f \partial \int \frac{\mathbf{m}(\mathbf{r}')}{|\mathbf{r}-\mathbf{r}'|}d^{3}\mathbf{r}'\right) ,
\end{eqnarray}
 whence we estimate
\[
m_{-}\sim (\omega -\omega _{L})m_{+}/2\omega _{L}\ll m_{+}.\]
 We see that in the vicinity of the Larmor frequency, when \( \omega \simeq \omega _{L} \),
the counter-rotating component \( m_{-} \) may be neglected with respect to
the co-rotating \( m_{+} \). Then (\ref{motion}) simplifies to a single linear
inhomogeneous integro-differential equation 
\begin{equation}
\label{schroedinger}
(\widehat{\mathcal{H}}-\omega )m_{+}(\mathbf{r})=\frac{\omega _{M}}{4\pi }h_{+}^{e}(\mathbf{r})
\end{equation}
 with a generally non-Hermitian Hamiltonian operator 
\begin{eqnarray}
\widehat{\mathcal{H}}m_{+} & = & D\left( 1+\ts {\frac{i}{C}}\right) \partial ^{2}m_{+}+\gamma H_{z}^{i}(\mathbf{r})m_{+}\nonumber \\
 & - & \frac{\omega _{M}}{8\pi }\partial _{+}\partial _{-}\int \frac{m_{+}(\mathbf{r}')}{|\mathbf{r}-\mathbf{r}'|}d^{3}\mathbf{r}'.
\end{eqnarray}
 We may use the equivalence \( \partial _{+}\partial _{-}=\partial ^{2}-\partial ^{2}_{z} \),
(\ref{HiM}) and the property (\ref{green-inf}) to rewrite the Hamiltonian
as
\begin{eqnarray}
\widehat{\mathcal{H}}m_{+} & = & D\left( 1+\ts {\frac{i}{C}}\right) \partial ^{2}m_{+}+\omega _{L}(\mathbf{r})m_{+}\nonumber \\
 & + & \frac{\omega _{M}}{2}\Bigl (\left( 1-2\underline{\widehat{n}}_{zz}[1]\right) m_{+}-\underline{\widehat{n}}_{zz}[m_{+}]\Bigr ).\label{dip-term} 
\end{eqnarray}

Here the integro-differential operator 
\[
\underline{\widehat{n}}_{zz}[f]=-\frac{1}{4\pi }\partial ^{2}_{z}\int _{V}\frac{f(\mathbf{r}')}{|\mathbf{r}-\mathbf{r}'|}d^{3}\mathbf{r}'\]
 is the \( zz \)-component of the demagnetizing tensor (\ref{demag-tensor}).
And \( \underline{\widehat{n}}_{zz}[1] \) is a scalar function on coordinates,
which actually coincides with the \( z \)-th demagnetizing coefficient \( n^{(z)} \)
for ellipsoids with one of the principal axes parallel to \( \hat{z} \). \( \omega _{L}(\mathbf{r}) \)
denotes \( \omega _{L}+z\nabla \omega _{L} \).

To conclude, we have derived equations of motion (\ref{motion}) or (\ref{schroedinger})
for small deviations of magnetization from static values. The magnetization
as a function of external rf field is a response of the system on a particular
radio frequency \( \omega  \). The full form of the equations of motion (\ref{motion})
is unnecessarily complicated because it contains superfluous information on
the dynamics of the counter-rotating component \( m^{-} \) of the magnetization.
An example of solution of the full Eq.~(\ref{motion}) for an infinite medium
in a uniform magnetic field is analyzed in Appendix \ref{AppInfinite}.

The rest of the paper deals with (\ref{schroedinger}).

The response of the liquid is detected through changes in the impedance of the
NMR coil, which are proportional to (see Appendix \ref{AppSpinSpectrum})
\begin{equation}
\label{self-inductance}
\overline{\chi }=\langle h^{e+}|\widehat{\mathcal{G}}_{\omega }|h^{e+}\rangle ,
\end{equation}
where \( \widehat{\mathcal{G}}_{\omega } \) is the Green operator 
\begin{equation}
\label{Green-operator}
|m^{+}\rangle =(\omega _{M}/4\pi )\widehat{\mathcal{G}}_{\omega }|h^{e+}\rangle .
\end{equation}

Green operator may be expanded into an infinite sum in the eigenfrequencies
\( \omega _{\alpha } \) of the homogeneous equation corresponding to (\ref{schroedinger})
\begin{equation}
\label{schroedinger1}
\widehat{\mathcal{H}}|\alpha \rangle =\omega |\alpha \rangle 
\end{equation}
 with the boundary condition
\begin{equation}
\label{boundary-condition}
\left. \widehat{n}_{i}\partial _{i}|\alpha \rangle \right| _{\partial }=0.
\end{equation}

For a Hermitian Hamiltonian the eigenfrequencies \( \omega _{\alpha } \) are
real and the expansion is 
\begin{equation}
\label{H-green}
\widehat{\mathcal{G}}_{\omega }=\sum _{\alpha }\frac{|\alpha \rangle \langle \alpha |}{\omega _{\alpha }-\omega }.
\end{equation}
 The absorption spectrum in the Hermitian case consists of a series of \( \delta  \)-peaks
at \( \omega =\omega _{\alpha } \). Indeed, writing real \( \omega _{\alpha } \)
in (\ref{H-green}) as \( \omega _{\alpha }+i0 \) we see that the imaginary
(absorption) part of (\ref{self-inductance}) is a weighted sum of \( \delta  \)-functions
\begin{equation}
\label{delta-set}
-\pi \sum _{\alpha }\left| \langle \alpha |h^{e+}\rangle \right| ^{2}\delta (\omega -\omega _{\alpha }).
\end{equation}

In the general case of a non-Hermitian Hamiltonian the expansion (\ref{H-green})
should be revised. We postpone the appropriate discussion until Sec. V. Here
it is enough to say that the numerators in the series (\ref{H-green}) remain
the same in the general case, but the eigenfrequencies \( \omega _{\alpha } \)
become complex, meaning that in general spectrum consists of Lorentzians.

The following important conclusion drawn on the basis of (\ref{H-green}) also
holds in the case of a non-Hermitian Hamiltonian. In a \emph{homogeneous} (\( \nabla \omega _{L}=0 \))
external static fields \( H^{e} \) and for ellipsoidal samples the Hamiltonian
(\ref{dip-term}) has uniform solutions, so-called Kittel modes \cite{kittel}
with the frequencies
\begin{equation}
\label{kittel-modes}
\omega _{0}=\omega _{L}+\frac{\omega _{M}}{2}\left( 1-3\underline{n}_{zz}\right) ,
\end{equation}
 where \( \underline{n}_{zz} \) is the \( \hat{z}\hat{z} \)-component of the
demagnetizing coefficient tensor \( \underline{n} \).

For customary sample sizes the rf field \( \mathbf{h}^{e} \) may be considered
spatially uniform. Then from (\ref{delta-set}) it follows that it is impossible
to excite a non-uniform mode by a homogeneous rf field \( \mathbf{h}^{e} \),
because then \( \langle \alpha |h^{e+}\rangle \propto \langle \alpha |0\rangle  \),
where \( |0\rangle  \) is the Kittel mode, and different modes are mutually
orthogonal \( \langle \alpha |0\rangle =\delta _{\alpha 0} \).

We conclude that in order to couple to non-uniform eigenmodes the external static
magnetic field should be \emph{inhomogeneous} (see (\ref{He})) so that there
would not exist a uniform eigenmode.

\section{Finite-cylindrical cell}

Study of the eigenstates of the Hamiltonian (\ref{dip-term}) in general is
possible only numerically. This has already been done in Ref.~\cite{candela 1},
\cite{candela 2} neglecting the contribution of the dipolar field. The former
work dealt with (\ref{dip-term}) in rectangular boxes, while the latter ---
in spherical containers.

In this paper we study the dipolar field effects due to the third term in (\ref{dip-term})
numerically in Section V. At the same time, in the case of \( \omega _{M}=0 \)
the problem allows explicit analytical solution for some typical experimental
conditions. Such solutions are of undeniable interest --- with them in hand
we may use perturbation theory to calculate corrections to modes in the first
order in \( \omega _{M} \). So the two following sections are dedicated to
such solutions and to the calculations of perturbational corrections respectively
in the geometries of a finite cylinder and a sphere.

We start the consideration of finite cylinders from idealized one-dimensional
geometry of a plane-parallel slab. Next we calculate the first order perturbations
to the modes frequencies due to the finiteness of the cylinder.

\subsection{Slab }

In the absence of dissipation, when \( C^{-1}=0 \), the effective diffusion
coefficient \( D(1+iC^{-1}) \) is real and thus the Hamiltonian \( \widehat{\mathcal{H}} \)
(\ref{dip-term}) is Hermitian.

In the slab geometry the solution should be sought in the form 
\[
m^{+}(\mathbf{r})\propto e^{i\mathbf{kr}_{\perp }}m_{\mathbf{k}}^{+}(z),\]
 where \textbf{\( \mathbf{r}_{\perp } \)} is the coordinate vector in the plane
perpendicular to \( \widehat{z} \). Nevertheless, as is clear from (\ref{self-inductance}),
only the solutions with \( \mathbf{k}=0 \) contribute to the observation signal
if the rf field \( \mathbf{h}^{e} \) is homogeneous.

The eigenfunctions \( m_{0}^{+}(z)=\langle z|\alpha \rangle  \) then are the
combinations of the two Airy functions
\begin{equation}
\label{1D-eigen}
\langle z|\alpha \rangle =A\mathrm{Ai}\left( \mathrm{arg}\right) +B\mathrm{Bi}\left( \mathrm{arg}\right) ,
\end{equation}
 where \( \mathrm{arg}=(\omega _{\alpha }-\omega _{L}-z\nabla \omega _{L})/\xi \nabla \omega _{L} \),
and
\begin{equation}
\label{lambda}
\xi =\sqrt[3]{D/\nabla \omega _{L}}
\end{equation}
 is the characteristic wavelength. Its sign depends on the relative sign of
\( D \) and \( \nabla \omega _{L} \) and thus on the sign of \( \kappa  \).
In \( ^{3} \)He and in \( ^{3} \)He-\( ^{4} \)He solutions with a concentration
\( x>3.5\% \), \( \kappa  \) is positive. We consider \( \xi >0 \) for definiteness.

The boundary conditions (\ref{boundary-condition}) on the two plane boundaries
\( \partial _{z}\left. |\alpha \rangle \right| _{z=0,L}=0 \) determine the
eigenfrequencies \( \omega _{\alpha } \) and the ratio of the coefficients
\( B/A \). The remaining coefficient \( A \) is determined from the normalization
condition \( \langle \alpha |\alpha \rangle =1 \).

When \( L\gg \xi  \) the influence of the lower wall of the container is negligible
and the modes are localized near the upper wall and decay exponentially into
the bulk on distances \( \sim \xi  \). Then the eigenfunctions are just the
Airy functions of the first kind --- \( \mathrm{Ai} \) and (\ref{1D-eigen})
becomes 
\[
\langle z|\alpha \rangle \equiv \langle z|n_{z}\rangle =A\mathrm{Ai}\left( \frac{L-z}{\xi }+\alpha '_{n_{z}}\right) ,\]
 where \( \alpha _{n}'<0 \) is the \( n \)-th zero of the derivative of the
Airy function \( \mathrm{Ai}' \): \( \alpha _{1}'\approx -1.02 \), \( \alpha _{2}'\approx -3.25 \),
\( \alpha _{3}'\approx -4.82 \), etc. The eigenfrequencies are 
\begin{equation}
\label{spectrum-slab}
\omega _{\alpha }\equiv \omega _{n_{z}}=\omega _{L}+L\nabla \omega _{L}+\alpha _{n_{z}}'\xi \nabla \omega _{L}.
\end{equation}

Inclusion of dissipation \( C^{-1}\neq 0 \) makes the diffusion coefficient
complex. The Hamiltonian (\ref{dip-term}) then becomes non-Hermitian. The complete
analysis of the spectra of a non-Hermitian Hamiltonian is possible only in the
framework of the general formalism to be developed in Sec. V.

However for the moment it is sufficient to make the following statement. In
the presence of dissipation (\( C^{-1}\neq 0 \)) \( \xi  \) in (\ref{lambda})
becomes complex 
\begin{equation}
\label{xi-substitute}
\xi \rightarrow \xi (1+i/C)^{1/3}
\end{equation}
 and so do arguments of the eigenfunctions (\ref{1D-eigen}). Eigenfrequencies
(\ref{spectrum-slab}) also acquire imaginary parts due to \( \xi  \) entering
the expression.

Thus the complex eigenfrequencies in the presence of dissipation can be easily
obtained from the real ones in the absence by the substitution (\ref{xi-substitute}).
This statement applies not only to (\ref{spectrum-slab}) but to any spectrum
of the Hamiltonian (\ref{dip-term}).

Let us now consider the effects of the dipolar field. We may utilize the results
of Appendix \ref{AppInfinite} since a slab is infinite in the direction perpendicular
to \( \hat{z} \) and modes depend on only \( z \). In such conditions the
demagnetizing field is local
\begin{equation}
\label{n-reduces-to-1}
\underline{\widehat{n}}_{zz}=n^{(z)}_{\mathrm{slab}}=1.
\end{equation}

It is then obvious from (\ref{dip-term}) that the dipolar field produces no
effect on the spin wave spectrum (\ref{spectrum-slab}) apart from a uniform
shift by 
\begin{equation}
\label{slab-shift}
\frac{1}{2}\omega _{M}(1-3n^{(z)}_{\mathrm{slab}})=-\omega _{M}.
\end{equation}

Such a shift does not distort the spectrum --- it does not change neither the
mutual positions of the modes nor their widths, from which the characteristics
of the liquid are derived.

We rather aim at finding those distortions, so we proceed to a more relevant
shape of a cylinder of a finite radius, which as well as all finite shapes as
we will see does give such distortions.

\subsection{Finite cylinder}

Consider a finite cylinder with the base radius \( R\gg \xi  \) and height
\( L\gg \xi  \) and a generatrix parallel to \( \hat{z} \). We find the influence
of the finiteness of a specimen on the magnitude of dipolar corrections to the
spectrum in the first order of perturbation theory.

The first order perturbational corrections to the modes frequencies are the
averages of the perturbation (dipolar) operator in the given eigenstate \( \psi _{\alpha }(\mathbf{r}) \)
\begin{eqnarray}
\delta _{\mathrm{dip}}\omega _{\alpha }=\frac{\omega _{M}}{2}\Bigl (1 & - & 2\int \left| \psi _{\alpha }(\mathbf{r})\right| ^{2}\underline{\widehat{n}}_{zz}[1]d^{3}\mathbf{r}\nonumber \label{ddip-cyl} \\
 & - & \int \psi ^{*}_{\alpha }(\mathbf{r})\underline{\widehat{n}}_{zz}[\psi _{\alpha }(\mathbf{r})]d^{3}\mathbf{r}\Bigr ).\label{ddip-cyl} 
\end{eqnarray}

In a finite cylinder the dipolar-free eigenfunction \( |\alpha \rangle  \)
satisfying boundary conditions (\ref{boundary-condition}), written in cylindrical
coordinates \( z,\rho ,\varphi  \) is 
\begin{equation}
\label{trhmodes}
\psi _{n_{z}n_{\rho }m}=c_{n_{z}n_{\rho }m}\mathrm{Ai}(\ts {\frac{L-z}{\xi }}+\alpha '_{n_{z}})J_{m}(\ts {\frac{\zeta _{n_{\rho }m}'}{R}}\rho )e^{im\varphi },
\end{equation}
 where \( n_{z},n_{\rho }=0,1,2,...,\infty  \) are respectively the longitudinal
and radial quantum numbers and \( m=-\infty ,...,+\infty  \) is the azimuthal
quantum number. \( \zeta _{n_{\rho }m}' \) is the (\( n_{\rho } \)+1)-th zero
of the derivative \( J'_{m} \) of the Bessel function \( J_{m} \), \( c_{n_{z}n_{\rho }m} \)
are the normalization coefficients.

From the general formula (\ref{self-inductance}) it is not hard to see that
only the modes with \( n_{\rho }=m=0 \) couple to the homogeneous rf field.
Indeed, \( |n_{z}00\rangle  \) is uniform in the plane perpendicular to \( \hat{z} \).
Therefore, integrals like \( \langle n_{z}n_{\rho }m|h^{+e}\rangle  \) are
proportional to \( \langle n_{\rho }m|00\rangle =\delta _{n_{\rho }0}\delta _{m0} \).

Calculating (\ref{ddip-cyl}) with (\ref{trhmodes}) yields (see Appendix \ref{AppDipFinCyl})

\begin{eqnarray}
\delta _{\mathrm{dip}}\omega ^{\mathrm{cylinder}}_{n_{z}} & = & \omega _{M}\Bigl (-1+\frac{L}{\pi R}\log \frac{8R}{eL}\label{slab-corrections} \\
 & + & \frac{\xi }{\pi R}\bigl (\Phi _{n_{z}}\log \frac{8R}{e^{2}\Xi _{n_{z}}\xi }+\Psi _{n_{z}}\log \frac{eL}{\Theta _{n_{z}}\xi }\bigr )\Bigr ).\nonumber 
\end{eqnarray}

The first two terms not dependent on the mode number \( n_{z} \) describe uniform
shift of the spectrum, and the last two terms dependent on \( n_{z} \) through
the numerical constants \( \Phi _{n_{z}} \), \( \Psi _{n_{z}} \), \( \Xi _{n_{z}} \)
and \( \Theta _{n_{z}} \), which are of the order of unity (see Table \ref{TableFinCyl}
in Appendix \ref{AppDipFinCyl}), give the sought-for spectrum distortion. 

We see that for finite \( \xi /R \) the spectrum undergoes distortion proportional
to the parameter 
\begin{equation}
\frac{2\omega _{M}}{\pi \xi \nabla \omega _{L}}\frac{\xi }{R}\log \frac{\sqrt{R^{\Phi }L^{\Psi }}}{\xi ^{\Phi +\Psi }},
\end{equation}
where \( \omega _{M}=4\pi \gamma M \) characterizes the magnetization density,
\( \nabla \omega _{L} \) is the gradient of the Larmor frequency, \( R \)
is the radius of the cylinder base, \( L \) its height and \( \xi  \) is the
wavelength (\ref{lambda}) of an Airy-type standing spin wave. The quantity
\( \xi \nabla \omega _{L} \) gives the average distance between modes in the
units of frequency. \( \Phi  \) and \( \Psi  \) are numbers of the order of
unity.

The calculations of the dipolar field effects in the first order of perturbation
theory allow us to estimate the maximum error due to the demagnetizing field
in the determination of the transverse relaxation time \( \tau  \) from the
spin wave spectra (see Appendix \ref{AppDipFinCyl}).

This error turns out to be of the order of the parameter (\ref{parameter}),
i.e. for the experiment \cite{vermeulen} \( \approx  \) 4.2\% for \( M/M_{0}\sim 4 \).

Thus, interpreting spectra according to the usual theory not taking into account
the demagnetizing field induces an error in the derived value of the transverse
relaxation time of the order of the parameter (\ref{parameter}).

The term proportional to \( \Phi _{n_{z}} \) comes from the demagnetizing field
produced by the rotating part \( \mathbf{m} \) of magnetization. While the
term proportional to \( \Psi _{n_{z}} \) is due to the spatial inhomogeneity
of the demagnetizing field \( -4\pi \underline{\widehat{n}}[\mathbf{M}] \)
produced by the initial static (homogeneous!) distribution of \( \mathbf{M} \)
(\ref{MHi}) in a finite cylinder.

\begin{figure}
\resizebox*{1\columnwidth}{!}{\includegraphics{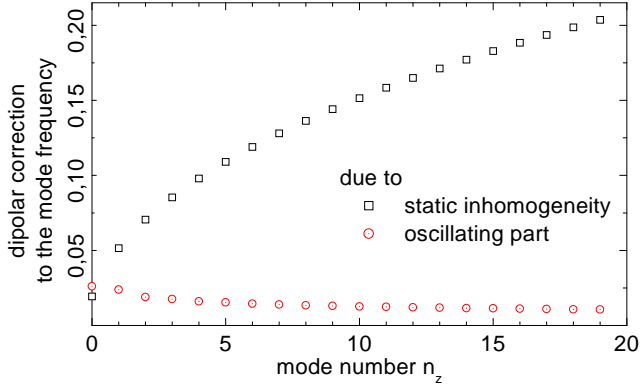}}

\caption{\label{Fig-CylTerms}Comparison of the two parts in the parameter of distortion
of the spin-wave spectrum in a finite-cylindrical cell brought about by the
demagnetizing field (the two last terms of (\ref{slab-corrections}) in units
of \protect\( \omega _{M}\protect \) for \protect\( \xi /R=0.015\protect \)
and \protect\( L=2R\protect \)). The first part (third term in (\ref{slab-corrections}))
(\protect\( \bigcirc \protect \)) is due to the demagnetizing field produced
by the rotating part \protect\( \mathbf{m}\protect \) of magnetization. The
second part (the last term in (\ref{slab-corrections})) (\protect\( \Box \protect \))
results from the inhomogeneity of the dipolar field produced by the initial
static distribution \protect\( \mathbf{M}=\chi _{n}AH^{e}\hat{z}\protect \)
of magnetization in a finite-cylindrical sample. }
\end{figure}

The values of these two terms are plotted in units of \( \omega _{M} \) as
functions of the mode number in Fig.~\ref{Fig-CylTerms} for \( \xi /R=0.015 \)
and \( L=2R \). Apart from being greater, the term brought about by the spatial
inhomogeneity of the static distribution of the dipolar field depends stronger
on the mode number, thus resulting in bigger mutual shifts of the modes. So
the main source of the spectrum distortion in a finite cylinder turns out to
be the inhomogeneity of the static dipolar field.

In ellipsoids, in particular in a sphere, the demagnetizing field produced by
the initial static distribution of \( \mathbf{M} \) is homogeneous. And so
there is only the distortion to the spectrum from the rotating part \( \mathbf{m} \)
of magnetization as we will see in the next section. This makes ellipsoidal
shapes advantageous if the dipolar field effects are unfavorable.

To conclude, we found the corrections to the spin wave modes in a finite cylinder
due to a weak demagnetizing field in perturbation theory. These corrections
consist in shifting the spectrum as a whole, changing the relative distances
between the modes and in narrowing down the modes. The two last are of interest
for us since they deform the spectrum.

There are two contributions to the spectrum deformation --- one from the static
inhomogeneous demagnetizing field and the other from the rotating part of the
magnetization. The first contribution exists only in non-ellipsoidal samples,
in which a homogeneous static magnetization produces inhomogeneous demagnetizing
field.

\section{Spherical cell}

Though for a spherical container exact analytical solution in the absence of
dipolar field turns to be impossible, one can obtain an explicit expression
for several first modes in adiabatic approximation if the radius of the sphere
\( R\gg \xi  \).

The prerequisites of adiabatic approximation might be best understood if one
exploits the analogy with the Schr\"{o}dinger equation for a particle moving
in an external field. If the movement in one direction is somehow more restricted
than in the others (geometrically or by an external field) it is a consequence
of Heisenberg uncertainty relations that the movement in this direction will
be faster. The slow enough movement in the other directions then will make up
an adiabatic perturbation that is known not to change the state of the particle
describing the fast motion.

As a result, the wave function can be combined as a multiplication of an envelope
depending only on the unrestricted coordinates, and of the fast motion state
depending on the unrestricted coordinates as on parameters.

\begin{figure}
\resizebox*{1\columnwidth}{!}{\includegraphics{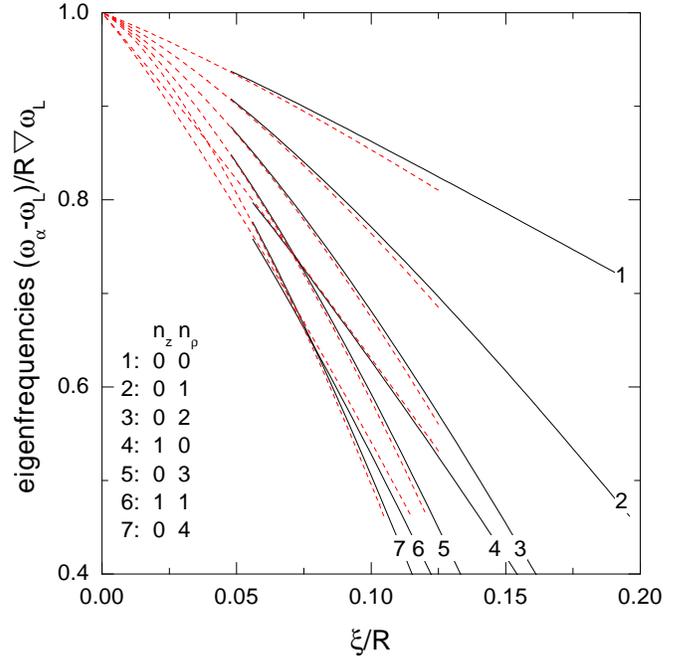}}

\caption{\label{Fig-comparison}Comparison of the first several spin wave modes frequencies
in a spherical cavity obtained numerically (solid curves, \cite{candela 2})
and in the adiabatic approximation (dashed curves, Eq.~(\ref{AD-f})). The eigenfrequencies
are plotted as functions of the ratio \protect\( \xi /R\protect \) of the characteristic
wavelength to the sphere radius. }
\end{figure}

This approach gives (see Appendix \ref{AppSph}) for the eigenfrequencies 
\begin{equation}
\label{spectrum-sphere}
\omega _{n_{z}n}=\omega _{L}+R\nabla \omega _{L}+\xi \nabla \omega _{L}\left[ \alpha _{n_{z}}'-\sqrt{2\frac{\xi }{R}}(2n_{\rho }+1)\right] ,
\end{equation}
 where \( \omega _{L} \) is the Larmor frequency in the center of the sphere,
and \( n_{z},n_{\rho }=0,1,2,...,\infty  \) are the longitudinal and radial
quantum numbers respectively. Since \( \xi /R\ll 1\sim \alpha _{n}' \) we see
that the lower-lying levels belong to \( \alpha _{0} \) and therefore decay
exponentially with diminishing \( z \). Modes with \( n_{z}=1 \) will make
a single oscillation before vanishing, modes with \( n_{z}=2 \) a double, etc.

Eq.~(\ref{spectrum-sphere}) reduces to (\ref{spectrum-slab}) in the limit
\( \xi /R\rightarrow 0 \) as it must. Indeed, in both (\ref{spectrum-sphere})
and (\ref{spectrum-slab}) there figures the Larmor frequency at the top of
a sample --- \( z=L \) for (\ref{spectrum-slab}) and \( z=R \) for (\ref{spectrum-sphere})
and the second term in brackets in (\ref{spectrum-sphere}) tends to zero when
\( \xi /R\rightarrow 0 \).

If we introduce dimensionless frequencies according to \cite{candela 2} \( \omega _{\alpha }=\omega _{L}+R\nabla \omega _{L}f_{\alpha } \),
we conclude that the observable eigenfrequencies for our problem are described
in the adiabatic approximation by 
\begin{equation}
\label{AD-f}
f_{\alpha }=1+\frac{\xi }{R}\left[ \alpha _{n_{z}}'-\sqrt{2\frac{\xi }{R}}(2n_{\rho }+1)\right] .
\end{equation}
 The comparison between this formula and the results obtained numerically in
Ref.~\cite{candela 2} is shown in Fig.~\ref{Fig-comparison}. In Ref.~\cite{candela 2}
the combination \( \xi /R \) was designated as \( \Delta  \).

Note that the numerical scheme developed in Ref.~\cite{candela 2} requires
more and more computational effort for \( \xi  \) tending to zero. The calculation
time to get safe eigenfrequencies values grows. That is why the numerical curves
are not shown in the vicinity of zero. Since we ourselves use a similar computational
technique we put off more detailed discussion until Section V.

Contrariwise, the discrepancy between approximate and numerical curves at large
\( \xi  \) is accounted for by inapplicability of adiabatics out of the region
\( \xi \ll R \).

So approximate and numerical methods complement each other. While numerics is
the method of choice for relatively large \( \xi /R \) when adiabatics breaks
down, it requires increasingly larger basis to obtain reliable results for small
\( \xi /R \). In this region it is easier to calculate eigenfrequencies in
adiabatic approximation.

\begin{figure}
\resizebox*{1\columnwidth}{!}{\includegraphics{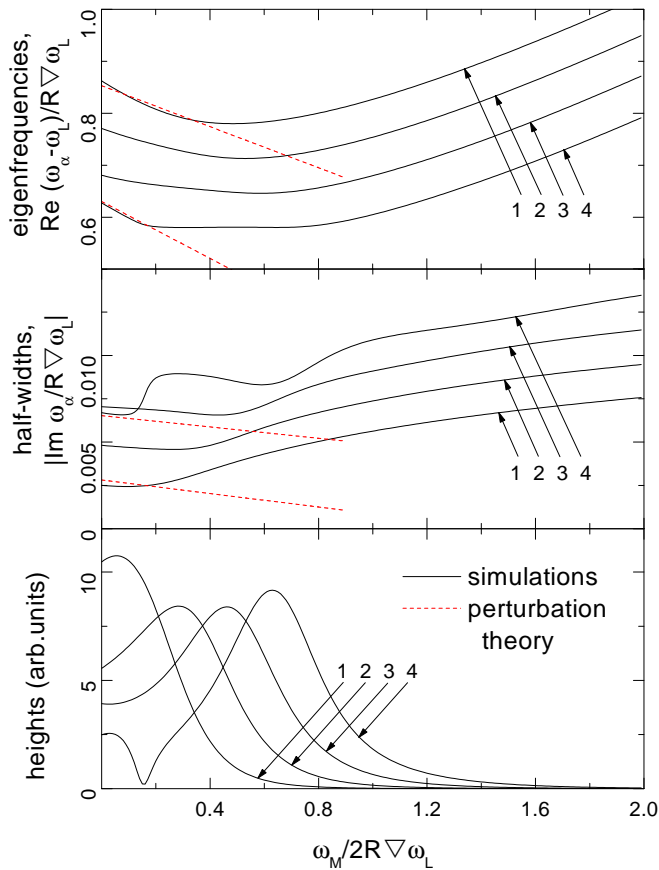}}

\caption{\label{Fig-modesOnOmega}The dependence of the frequencies, half-widths and
heights of the first four modes (numbered in the same order as on Fig. 2) on
the strength of the demagnetizing field \protect\( \omega _{M}\protect \) (solid
curves) obtained by numerical simulations in Section V. Input parameters are
\protect\( \xi /R=0.1\protect \), \protect\( C=20\protect \). First order
perturbational corrections (dashed lines) calculated in the text belong to the
region \protect\( \omega _{M}/2R\nabla \omega _{L}\ll 1\protect \). Perturbational
calculations were done only for the modes with radial quantum numbers \protect\( n_{\rho }=0\protect \),
among the four depicted only the first and the fourth are of the type. Heights
of the modes are impossible to calculate perturbationally.}
\end{figure}

Let us now look on dipolar field correction to the modes.

Quite analogously to the case of a finite cylinder, we obtain for the corrections
to the modes (see Appendix \ref{AppSph}) 
\begin{equation}
\label{sphere-corrections}
\delta _{\mathrm{dip}}\omega ^{\mathrm{sphere}}_{\alpha }=\omega _{M}\left( -\frac{1}{3}+\frac{\Phi _{n_{z}}\sqrt{\pi }}{4}\sqrt[4]{\frac{\xi }{2R}}\right) .
\end{equation}
 And correspondingly for the corrections to the imaginary parts of the modes
\begin{equation}
\delta _{\mathrm{dip}}\Im \omega ^{\mathrm{sphere}}_{\alpha }=\frac{\omega _{M}}{12C}\frac{\Phi _{n_{z}}\sqrt{\pi }}{4}\sqrt[4]{\frac{\xi }{2R}}.
\end{equation}
 Numerical constants \( \Phi _{n_{z}} \) are the same as in (\ref{slab-corrections}). 

The parameter determining the relative value of the dipolar field effects in
a sphere is
\begin{equation}
\label{parameter-sph-text}
\frac{\sqrt{\pi }\omega _{M}}{4\xi \nabla \omega _{L}}\sqrt[4]{\frac{\xi }{2R}}.
\end{equation}

In the next section we will solve the eigenvalue problem in a sphere in the
presence of a dipolar field of an arbitrary strength. It is interesting to compare
the results of numerical simulations with the first order perturbational corrections
written above. 

\begin{figure}
\resizebox*{1\columnwidth}{!}{\includegraphics{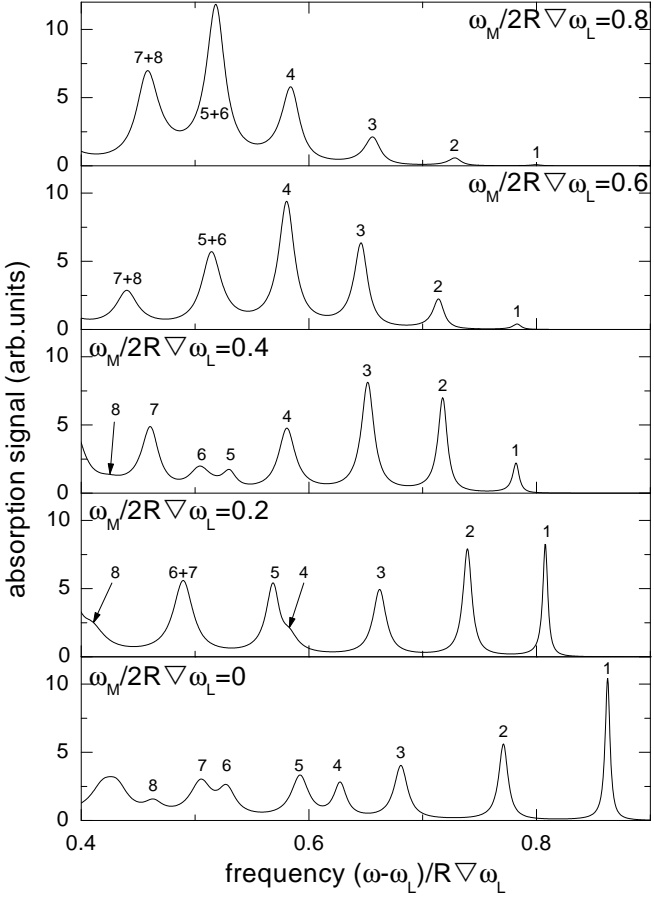}}

\caption{\label{Fig-compareModes0.eps} Spin wave spectra obtained numerically in Section
V for several \protect\( \omega _{M}/2R\nabla \omega _{L}<1\protect \). Input
parameters are \protect\( \xi /R=0.1\protect \), \protect\( C=20\protect \).
Modes are numbered in the same order as on Fig. 2. }
\end{figure}

To this end the positions \( \Re (\omega _{\alpha }-\omega _{L}) \) and half-widths
\( \Im \omega _{\alpha } \) of the first four modes obtained from both numerics
and analytics are plotted on Fig.~\ref{Fig-modesOnOmega} for several \( \omega _{M} \).
Heights are plotted only as obtained numerically. As one can see from Fig. \ref{Fig-modesOnOmega},
perturbational approximation is satisfactory for the values of \( \omega _{M}/2R\nabla \omega _{L} \)
up to \( \sim 0.1 \), or, for values of (\ref{parameter-sph-text}) up to \( \sim 0.4 \).
The small discrepancy between numerical and analytical results even in this
region is accounted for by the restrictions of the adiabatic approximation used
to fulfill analytical calculations.

The spectra themselves calculated numerically for several values of \( \omega _{M} \)
are shown on Fig.~\ref{Fig-compareModes0.eps}. Modes weights are redistributing
between adjacent modes with \( \omega _{M} \) growing. Not only the weights
but also half-widths and positions of the modes change.

\section{Sphere. Numerical calculations}

The complexity of the Hamiltonian (\ref{dip-term}) does not allow finding exact
eigenfunctions by analytical methods apart from using perturbation theory. Nevertheless
it is always possible to solve equation (\ref{schroedinger1}) numerically.
We chose a spherical container for numerical investigations.

Before proceeding to the description of the simulation scheme utilized, a formalism
is to be established for solving an eigenvalue problem with a non-Hermitian
Hamiltonian. Of interest for us is the generalization of the expansion (\ref{H-green})
of the Green function into series over eigenfunctions. The following subsection
is dedicated to the topic.

In the two remaining subsections we discuss respectively the technique and the
results, other than already mentioned in the previous section, of the numerical
simulations.

\subsection{Green function of a non-Hermitian Hamiltonian}

In this subsection we derive an analogue of (\ref{H-green}) for a non-Hermitian
Hamiltonian \( \widehat{\mathcal{H}}\neq \widehat{\mathcal{H}}^{+} \). For
\( \widehat{\mathcal{H}}\neq \widehat{\mathcal{H}}^{+} \) the set of eigenfunctions
\( |\alpha \rangle  \) of \( \widehat{\mathcal{H}} \) is not orthogonal. A
second set of functions, viz., the set of eigenfunctions of the Hermitian conjugate
operator \( \widehat{\mathcal{H}}^{+} \), is to be introduced \cite{morse}.

The eigenvalues of the operator \( \widehat{\mathcal{H}}^{+} \) are just the
complex conjugates of the eigenvalues of \( \widehat{\mathcal{H}} \). Indeed,
the eigenvalues are found from a secular equation. And for the Hermitian conjugate
operators these algebraic equations may be shown to be complex conjugate.

Thus the set of eigenfunctions of \( \widehat{\mathcal{H}}^{+} \) may always
be numbered with the same index \( \alpha  \). To distinguish this set from
\( |\alpha \rangle  \) we will denote it by \( |\alpha \rev  \)
\begin{eqnarray}
\widehat{\mathcal{H}}|\alpha \rangle  & = & \omega _{\alpha }|\alpha \rangle ,\\
\widehat{\mathcal{H}}^{+}|\alpha \rev  & = & \omega ^{*}_{\alpha }|\alpha \rev .\label{H+} 
\end{eqnarray}
 However, it should be borne in mind that \( |\alpha \rev  \) like \( |\alpha \rangle  \)
is an ordinary set of ket-vectors, which has the corresponding set of bra-vectors.

It turns out then that, notwithstanding that neither the set \( |\alpha \rangle  \)
nor \( |\alpha \rev  \) is orthogonal, there is orthogonality \emph{between
the two sets.} Indeed, following a conventional scheme of proving mutual orthogonality
of eigenfunctions, we note that 
\[
\lev \alpha |\widehat{\mathcal{H}}|\beta \rangle =\omega _{\beta }\lev \alpha |\beta \rangle .\]
 On the other hand, 
\begin{equation}
\lev \alpha |\widehat{\mathcal{H}}=\left( \widehat{\mathcal{H}}^{+}|\alpha \rev \right) ^{+}=\left( \omega ^{*}_{\alpha }|\alpha \rev \right) ^{+}=\omega _{\alpha }\lev \alpha |.
\end{equation}
 Multiplying this by \( |\beta \rangle  \) and subtracting the previous result,
we see that
\begin{equation}
(\omega _{\alpha }-\omega _{\beta })\lev \alpha |\beta \rangle =0.
\end{equation}

Thus \( |\beta \rangle  \) and \( |\alpha \rev  \) are orthogonal if \( \alpha \neq \beta  \).
So it is said that the two sets \( |\alpha \rangle  \) and \( |\alpha \rev  \)
constitute a \emph{biorthogonal set} of eigenfunctions.

The expansion of an arbitrary function into a convergent series is then possible
\begin{equation}
|x\rangle =\sum _{\alpha }|\alpha \rangle \lev \alpha |x\rangle .
\end{equation}
 Here the eigenfunctions are supposed to be normalized so that 
\begin{equation}
\lev \alpha |\beta \rangle =\delta _{\alpha \beta }.
\end{equation}

Performing such an expansion for \( |h^{+}\rangle  \) in (\ref{schroedinger}),
one obtains 
\begin{equation}
\label{nonH-green}
\widehat{\mathcal{G}}_{\omega }=\sum _{\alpha }\frac{|\alpha \rangle \lev \alpha |}{\omega _{\alpha }-\omega }.
\end{equation}
 This is the sought-for generalization of (\ref{H-green}). For a Hermitian
Hamiltonian the two sets coincide \( |\alpha \rangle =|\alpha \rev  \), and
(\ref{nonH-green}) reduces to (\ref{H-green}).

Thus we see from (\ref{self-inductance}) that the spectrum of a non-Hermitian
Hamiltonian consists of a set of Lorentzians at \( \omega =\Re \omega _{\alpha } \)
with half-widths \( \Im \omega _{\alpha } \) each entering with a weight \( \langle h^{e+}|\alpha \rangle \lev \alpha |h^{e+}\rangle  \).
If the rf field can be regarded as uniform on the scales of the sample, the
relative weights of the Lorentzian peaks are
\begin{equation}
\label{modes-weights}
\int \langle \mathbf{r}|\alpha \rangle d^{3}\mathbf{r}\int \lev \alpha |\mathbf{r}'\rangle d^{3}\mathbf{r}'.
\end{equation}

In our particular case, the Hamiltonian (\ref{dip-term}) is symmetric \( \widehat{\mathcal{H}}^{+}=\widehat{\mathcal{H}}^{*} \).
This is trivial to see without the dipolar field term, but this term can also
be shown to be (real) symmetric, since partial differential operator \( \partial ^{2}_{z} \)
and the Green operator \( \hat{\mathcal{G}}_{\infty } \) are both real symmetric
and commutative \( \partial ^{2}_{z}\hat{\mathcal{G}}_{\infty }=\hat{\mathcal{G}}_{\infty }\partial ^{2}_{z} \)
(We remind that for an integral operator with a kernel \( G(\mathbf{r},\mathbf{r}') \)
Hermitian conjugate has the kernel \( G^{*}(\mathbf{r}',\mathbf{r}) \)). Indeed,
we write 
\begin{equation}
\partial ^{2}_{z}\hat{\mathcal{G}}_{\infty }\mathbf{m}=\partial ^{2}_{z}\int \frac{\mathbf{m}(\mathbf{r}')}{|\mathbf{r}-\mathbf{r}'|}d^{3}\mathbf{r}'=\int \mathbf{m}(\mathbf{r}')\partial ^{2}_{z'}\frac{1}{|\mathbf{r}-\mathbf{r}'|}d^{3}\mathbf{r}'.
\end{equation}
 Taking the integral two times by parts and taking into account that \( \mathbf{m}(\mathbf{r}')/|\mathbf{r}-\mathbf{r}'|\rightarrow 0 \)
at \( z'\rightarrow \pm \infty  \), we get 
\begin{equation}
\int \frac{\partial ^{2}_{z'}\mathbf{m}(\mathbf{r}')}{|\mathbf{r}-\mathbf{r}'|}d^{3}\mathbf{r}'=\hat{\mathcal{G}}_{\infty }\partial ^{2}_{z}\mathbf{m}.
\end{equation}

Therefore from the definition (\ref{H+}) we immediately conclude that in the
case of a symmetric Hamiltonian \( \widehat{\mathcal{H}}^{+}=\widehat{\mathcal{H}}^{*} \)
the two sets of eigenfunctions are related through \( \langle \mathbf{r}|\alpha \rev =\langle \mathbf{r}|\alpha \rangle ^{*}\equiv \langle \alpha |\mathbf{r}\rangle  \).
Expression (\ref{modes-weights}) for the relative weights of the modes simplifies
then to 
\begin{equation}
\label{Ham-symm-modes-weights}
\left( \int \langle \mathbf{r}|\alpha \rangle d^{3}\mathbf{r}\right) ^{2}.
\end{equation}
 Note that the (complex) value itself of the integral is squared, not its absolute
value, as it would be should the Hamiltonian be Hermitian. We will use the expression
(\ref{Ham-symm-modes-weights}) for the modes weights in numerical calculations
of the spectra.

\subsection{Numerical approach}

One of the methods for solving a spectral Sturm-Liouville problem (conceptually,
perhaps, the simplest) consists in its finite-dimensional approximation. Formally,
we then are left with standard algebraic spectral problem. For the solution
of the latter one can implement one of ready safe well-established algorithms.
However, one must be cautious with the dimension of approximation.

One of the possible discretization techniques is to find eigenfunctions in the
representation of some complete orthonormal set of functions when the Hamiltonian
(\ref{dip-term}) would become a matrix. Such a scheme was developed in application
to spin waves in Ref.~\cite{candela 1}, \cite{candela 2}.

A handy orthonormal set to choose is that of eigenfunctions of the Laplace operator
satisfying the boundary conditions for the geometry in question 
\begin{eqnarray}
\left[ \partial ^{2}+k_{\mu }^{2}\right] |\mu \rangle  & = & 0,\label{eigenLaplace} \\
\left. \widehat{n}_{i}\partial _{i}|\mu \rangle \right| _{\partial } & = & 0.\label{bcLaplace} 
\end{eqnarray}
 Here \( \mu  \) stands for a complete set of indices needed to describe a
state, \( k_{\mu } \) are waveconstants. In the case of such a choice of the
set the first term of the Hamiltonian (\ref{dip-term}) becomes trivial and
the boundary conditions are met automatically.

The eigenfunctions \( |\alpha \rangle  \), \( |\alpha \rev  \) of the operators
\( \widehat{\mathcal{H}} \), \( \widehat{\mathcal{H}}^{+} \) then take the
form 
\begin{equation}
|\alpha \rangle =\sum _{\mu }|\mu \rangle \langle \mu |\alpha \rangle ,\quad \quad |\alpha \rev =\sum _{\nu }|\nu \rangle \langle \nu |\alpha \rev .
\end{equation}
 where the coefficients of expansion \( \langle \mu |\alpha \rangle  \) and
\( \langle \nu |\alpha \rev  \) are found numerically as right and left eigenvectors
of the matrix form of the Hamiltonian (\ref{dip-term}) corresponding to eigenfrequencies
\( \omega _{\alpha } \)

\begin{eqnarray}
\sum _{\nu }\mathcal{H}_{\mu \nu }\langle \nu |\alpha \rangle  & = & \omega _{\alpha }\langle \mu |\alpha \rangle ,\label{matrix-H-right} \\
\sum _{\nu }\lev \alpha |\nu \rangle \mathcal{H}_{\nu \mu } & = & \omega _{\alpha }\lev \alpha |\mu \rangle ,\label{matrix-H-left} 
\end{eqnarray}
 where \( \mathcal{H}_{\mu \nu }=\langle \mu |\widehat{\mathcal{H}}|\nu \rangle  \).

For a sphere \( \mu  \) denotes the set \( n \), \( l \), \( m \) of the
radial, polar and azimuthal quantum numbers: \( n \), \( l=0 \), \( 1 \),
\( 2 \), \( \ldots  \), \( m=-l \), \( -l+1 \), \( \ldots  \), \( l-1 \),
\( l \). The corresponding basis is 
\begin{equation}
\label{mu}
\langle \mathbf{r}|\mu \rangle =\langle \mathbf{r}|nlm\rangle =c_{nl}j_{l}(k_{nl}r)Y^{m}_{l}(\widehat{\mathbf{r}})
\end{equation}
 where \( j_{l} \) is the spherical Bessel function, \( Y^{m}_{l} \) is the
spherical harmonic, and renormalization coefficients \( c_{nl} \) are defined
according to 
\begin{equation}
\label{sph-norm}
c_{nl}c_{n'l}\int ^{R}_{0}j_{l}(k_{nl}r)j_{l}(k_{n'l}r)r^{2}dr=\delta _{nn'}.
\end{equation}

The waveconstants \( k_{nl} \) depend on the boundary condition. That of (\ref{bcLaplace})
requires that \( k_{nl}R \) be the \( (n+1) \)-th zero of the derivative \( \partial _{r}j_{l}(r) \)
of the spherical Bessel function.

Since we are interested only in axisymmetric modes, which couple to a homogeneous
rf field, we may simplify the formulas by working with a sub-basis \( |nl0\rangle  \).

Infinite indexing \( n \), \( l \) is to be truncated at some finite values
for numerical computation. Maximum values of \( n_{\mathrm{max}} \), \( l_{\mathrm{max}} \)
are restricted by computational tractability of resulting matrices.

On the other hand, justification for such a truncation is that coefficients
\( \langle nl|\alpha \rangle  \) tend to zero for large \( n \), \( l \)
because of the oscillating character of \( j_{l} \). We expect \( \langle nl|\alpha \rangle  \)
close to zero when the characteristic scale \( \xi  \) of the function \( \langle \mathbf{r}|\alpha \rangle  \)
becomes greater then the period \( \sim R/n \) of the oscillations of the basis
function \( j_{l} \). Empirically, \( n_{\mathrm{max}}=10 \) is already quite
good for customary \( \xi /R\sim 0.1 \). Note that the change in \( j_{l}(k_{nl}r) \)
with increasing \( l \) is much less dramatic. So more \( l \)'s are to be
retained in the sub-basis. We used \( l_{\mathrm{max}}=51 \). Further increase
of \( n_{\mathrm{max}} \), \( l_{\mathrm{max}} \) proved to have no apparent
effect on the spectra for \( \xi /R\sim 0.1 \).

However, for \( \xi /R\rightarrow 0 \) more and more sub-basis functions should
be kept, which leads to rapid slowing down of the computations. In this limit
adiabatic approximation (see Sec.~IV) gives safer results.

It is convenient to seek for the eigenfrequencies \( \omega _{\alpha } \) in
the form \( \omega _{L}+R\nabla \omega _{L}f_{\alpha } \). The Hamiltonian
for the matrix equation on \( f_{\alpha } \)
\begin{eqnarray}
\langle nl|\widehat{\mathcal{H}}_{f}|n'l'\rangle  & = & -\ts {\bigl (\frac{\xi }{R}\bigr )^{3}\left( 1+\frac{i}{C}\right) }\left( k_{nl}R\right) ^{2}\delta _{nn'}\delta _{ll'}\nonumber \\
 & + & \left\langle nl\left| \ts {\frac{z}{R}}\right| n'l'\right\rangle \label{f-ham} \\
 & + & \ts {\frac{\omega _{M}}{2R\nabla \omega _{L}}}\left\langle nl\left| \ts {\frac{1}{3}}-\underline{\widehat{n}}_{zz}\right| n'l'\right\rangle \nonumber 
\end{eqnarray}
 comprises three parameters --- \( \xi /R \), the ratio \( \omega _{M}/2R\nabla \omega _{L} \)
of \( \omega _{M} \) to the total field gradient over the sample and the regime
parameter \( C \). Here we have used \( 1-2n_{\mathrm{sphere}}^{(z)}=\frac{1}{3} \).

The matrix elements of \( z/R \) and of \( \underline{\widehat{n}}_{zz} \)
by integrating over the solid angle with appropriate spherical functions reduce
to integrals over the radial coordinate \( r \).

The matrix elements of \( z/R \) were written in \cite{candela 2}, the integrals
arising should be calculated numerically. The rather lengthy calculations of
the matrix elements of \( \underline{\widehat{n}}_{zz} \) were separated into
Appendix \ref{AppMatrixElements}.

We cite here only the results. The matrix elements of \( z/R=r\cos \theta /R \),
where \( \theta  \) is the spherical polar angle are non-zero only if \( l'=l\pm 1 \)
\begin{eqnarray}
\langle nl| & z/R & |n',l'=l\pm 1\rangle \label{grad-me} \\
 & = & c^{0}_{\lambda }c_{nl}c_{n'l'}\int ^{R}_{0}j_{l}(k_{nl}r)j_{l'}(k_{n'l'}r)r^{3}dr/R.\nonumber 
\end{eqnarray}
 Here \( \lambda  \) is the greater of \( l \), \( l' \) and
\begin{equation}
c_{\lambda }^{0}=\lambda /\sqrt{4\lambda ^{2}-1}.
\end{equation}

The matrix elements of \( \underline{\widehat{n}}_{zz} \) are non-zero only
if \( l'=\{l,l\pm 2\} \) (see (\ref{App-me}) in Appendix \ref{AppMatrixElements})
\begin{eqnarray}
 & \langle n & l|\ts {\frac{1}{3}}-\underline{\widehat{n}}_{zz}|n'l\rangle =\delta _{nn'}\left[ \ts {\frac{1}{3}}-\left( c_{l}^{0}\right) ^{2}-\left( c_{l+1}^{0}\right) ^{2}\right] ,\label{me-dip} \\
 & \langle n & l|\ts {\frac{1}{3}}-\underline{\widehat{n}}_{zz}|n',l'=l\pm 2\rangle =c^{0}_{\lambda }c^{0}_{\lambda -1}c_{nl}c_{n'l'}R^{2}\nonumber \\
 & \times  & \frac{k_{nl}j_{l+1}(k_{nl}R)j_{l}(k_{n'l'}R)-k_{n'l'}j_{l+1}(k_{n'l'}R)j_{l}(k_{nl}R)}{k_{nl}^{2}-k_{n'l'}^{2}}\nonumber 
\end{eqnarray}

It may be verified that the matrix \( \langle nl|\widehat{\mathcal{H}}_{f}|n'l'\rangle  \)
is indeed symmetric.

The algebraic eigenvalue problem for (\ref{f-ham}) with (\ref{grad-me}) was
solved using standard subroutine from the Linear Algebra Package LAPACK. Eigenvectors
were then normalized and the left and right eigenvectors used to find the modes
weights.

In calculating modes weights using (\ref{matrix-H-right}), (\ref{matrix-H-left})
we note that \( \langle \mathbf{r}|000\rangle =1/\sqrt{V} \), where \( V=4\pi R^{3}/3 \)
is the sphere volume. Thus 
\begin{equation}
\int \langle \mathbf{r}|nlm\rangle d^{3}\mathbf{r}=\sqrt{V}\langle 000|nlm\rangle =\sqrt{V}\delta _{n0}\delta _{l0}\delta _{m0}.
\end{equation}
 The general formula (\ref{modes-weights}) in this case reduces to 
\begin{equation}
V\langle 000|\alpha \rangle \lev \alpha |000\rangle .
\end{equation}

\subsection{Results of simulations}

\begin{figure}
\resizebox*{1\columnwidth}{!}{\includegraphics{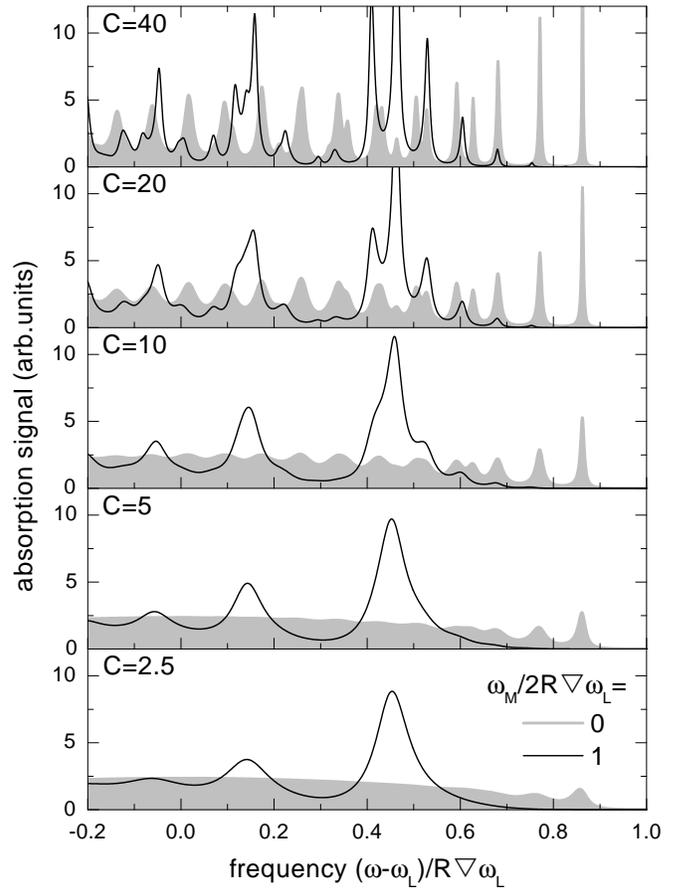}}

\caption{\label{Fig-o0o1onC} Absorption signals of a polarized Fermi liquid in a spherical
container for several regime parameters \protect\( C\protect \) in neglect
of the dipolar field effects \protect\( \omega _{M}=0\protect \) (grey filled
curves) and for a dipolar field of an intermediate strength \protect\( \omega _{M}/2R\nabla \omega _{L}=1\protect \)
(solid curves). The ratio \protect\( \xi /R\protect \) was put \protect\( 0.1\protect \). }
\end{figure}

Numerical results in the absence of the dipolar field were obtained in Ref.~\cite{candela 2}.
When \( \omega _{M}=0 \) there remain only two parameters in the problem ---
the ratio \( \xi /R \) of the characteristic wavelength to the sphere radius
and the regime parameter \( C \) which determines half-widths of the modes.
The dependence of the spin wave spectrum on \( \xi /R \) in the limit of weak
demagnetizing field was plotted in Fig.~\ref{Fig-comparison} to compare adiabatic
approximation with simulations. A typical absorption signal in the absence of
the dipolar field is depicted as grey filled curves in Fig.~\ref{Fig-o0o1onC}
for several \( C \). Decrease in \( C \) results in modes broadening without
changing their positions.

When \( \omega _{M} \) is small enough the dipolar field constitutes a perturbation
to conventional Silin spin waves. It is this regime which was studied perturbationally
in Sec. IV.2.

In this subsection we shortly consider other results of numerical calculations,
viz., regimes of intermediate and strong demagnetizing fields. Although these
regimes were not realized so far in Fermi liquids (see Table I), one cannot
leave out what is to be expected.

\begin{figure}
\resizebox*{1\columnwidth}{!}{\includegraphics{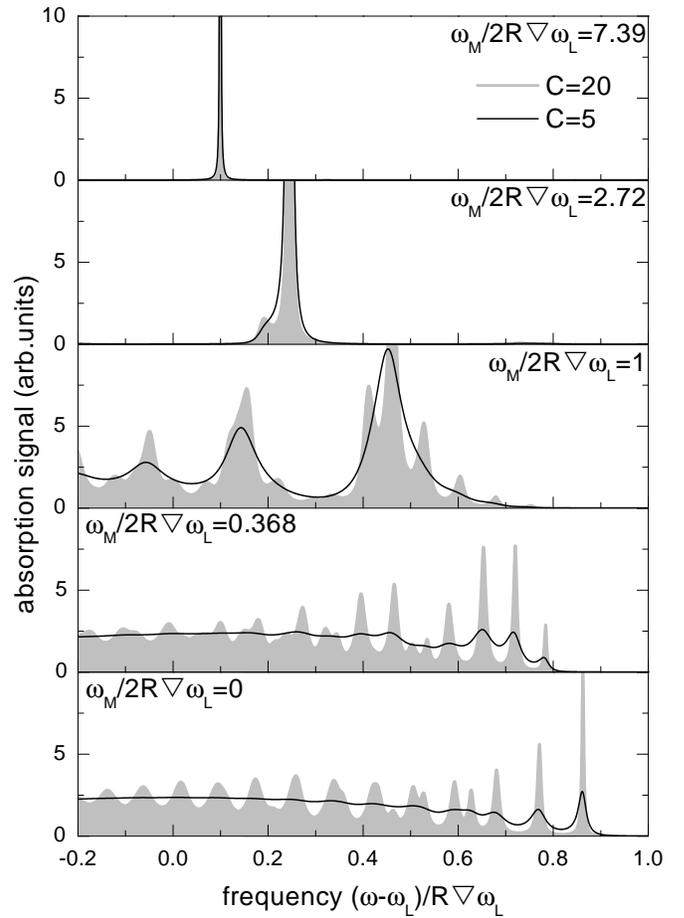}}

\caption{\label{Fig-C5C20onO} From bottom to top --- evolution of the spin waves spectrum
of a polarized Fermi liquid at a crossover from weak to strong demagnetizing
field. Several adjacent Silin modes form a group which evolves further as a
magnetostatic whole. The distance between magnetostatic modes grows with \protect\( \omega _{M}\protect \)
until we are left with the sole mode in which magnetization in the whole sample
oscillates uniformly. Input parameters were \protect\( \xi /R=0.1\protect \),
\protect\( C=20\protect \) (grey filled curves) and \protect\( C=5\protect \)
(solid curves). }
\end{figure}

As \( \omega _{M} \) increases the spin wave spectrum undergoes crossover from
Silin type for small demagnetizing fields (\( \omega _{M}/2R\nabla \omega _{L}<1 \))
to magnetostatic type for large demagnetizing fields (\( \omega _{M}/2R\nabla \omega _{L}>1 \)).
On a gross scale this transition is represented in Fig.~\ref{Fig-C5C20onO}
for \( \xi /R=0.1 \) and \( C=20 \) (grey filled curves) and \( C=5 \) (solid
curves). Modes weights change so that adjacent Silin modes group into fewer
magnetostatic modes. These separate at even larger \( \omega _{M} \) until
a uniform (Kittel) mode singles out at extremely large \( \omega _{M} \).

This latter mode is the only one to remain because we chose a uniform radio-frequency
field for the response of the system. And a non-uniform rf field is required
to couple to non-uniform magnetostatic modes since the influence of external
field gradient \( \nabla \omega _{L} \) is negligible for large \( \omega _{M} \).

The behavior described is not altered by larger dissipation (smaller \( C \))
other than Silin modes grouping becomes more pronounced (see solid curves on
Fig.~\ref{Fig-C5C20onO}).

The dependence on \( C \) of a spectrum for a demagnetizing field of an intermediate
strength (\( \omega _{M}/2R\nabla \omega _{L}=1 \)) is plotted in Fig.~\ref{Fig-o0o1onC}
as solid curves. Pronounced adjacent modes for larger \( C \) merge into magnetostatic
conglomerates with no distinction for smaller \( C \). No apparent relation,
especially for smaller \( C \), can be seen with the spectrum in the absence
of the demagnetizing field.

\section{Discussions}

\subsection{The method}

We studied the influence of the dipolar (or demagnetizing) field on the spectrum
of linear standing spin waves in a polarized Fermi liquid in a finite container.

A somewhat resembling problem was studied in the 70s in ferrimagnets. deWames
and Wolfram \cite{patton} considered not the Larmor-precession case (\ref{precession}),
but the true Landau-Lifshitz equations of inhomogeneous magnetization dynamics
that contain an additional exchange torque term depending on the Laplacian of
the magnetization. This term is similar to that arising from the Leggett equations
(see Eq.~(\ref{torque-term})).

The situation in ferrimagnets and in polarized Fermi-liquids belong however
to different limiting cases. In ferrimagnets the dipolar field effects --- magnetostatic
waves --- are pronounced and exchange is a perturbation. In a paramagnetic Fermi
liquid under conventional experimental conditions, on the contrary, it is the
dipolar field whose effect is smaller.

There is another obstacle of transplanting the approach of deWames and Wolfram
to our needs. In order to accommodate all the inhomogeneities --- both from
the Landau-Lifshitz equations and from the Maxwell equations (\ref{ms}) ---
the resulting differential equation for magnetostatic potential is of the sixth
order in spatial derivatives. Though it was manageable in a quasi-1D situation
(ferrimagnetic substance yttrium iron garnet YIG is grown as thin films), it
becomes virtually intractable in any, even the simplest, 3D geometry.

An approach to include dipolar field into magnetization dynamics analogous to
ours was used by Deville et al. \cite{Deville} in application to solid bcc
\( ^{3} \)He. In this material the dipolar field gives rise to multiple spin
echos at times \( n\tau  \) following two isolated rf pulses at times \( 0 \)
and \( \tau  \). Ref.~\cite{Deville} explained this phenomenon quantitatively
using a simplified local approximation for the dipolar field valid for a slab
infinite in the directions perpendicular to the external field (although this
restriction was not properly emphasized in the paper). Later Fomin and Vermeulen
\cite{FominVermeulen} utilized the form of the dipolar field term of Ref.~\cite{Deville}
to study dipolar corrections to a two-domain coherently precessing structure.

As for linear spin waves, in this local approximation neither the relative positions
of the modes nor their half-widths and heights change, and the spectrum only
shifts as a whole. The demagnetizing field is by definition shape-dependent
and it is erroneous to use the local approximation \emph{a priori}.

\subsection{Main results}

The strength of the demagnetizing field is proportional to the polarization
of the liquid. So in conventional weakly-polarized liquids the dipolar field
can be neglected, while for strong enough polarizations its effects overwhelm
usual Fermi liquid exchange field spin dynamics.

Because of its long-range character, the dipolar interaction introduces an additional
non-local term into the equations of the spin dynamics. This term is an integro-differential
operator on the magnetization, wherein the integration is taken over the volume
of the liquid. As a consequence the corrections to the spin dynamics depend
strongly on the shape of the container.

\begin{widetext}

\begin{table}

\caption{\label{conditions}Comparison of the conditions of different spin-wave experiments
\cite{vermeulen}, \cite{candela 1}, \cite{candela 2}, \cite{Roni} in pure
\protect\( ^{3}\protect \)He and in solutions of \protect\( ^{3}\protect \)He
in \protect\( ^{4}\protect \)He. From left to right. i) The concentration \protect\( x\protect \)
of \protect\( ^{3}\protect \)He atoms in mixtures (for pure \protect\( ^{3}\protect \)He
not given). ii) The static magnetization \protect\( M\protect \) in the units
of the equilibrium magnetization \protect\( M_{0}\protect \) (\ref{equilibrium-mag})
in the external field; different from 1 only in \cite{vermeulen}, \cite{Roni};
see the text for more explanation. iii) The Larmor frequency \protect\( \omega _{L}/2\pi \protect \),
where \protect\( \omega _{L}=\gamma H^{e}\protect \). iv) The gradient \protect\( \nabla H^{e}=\nabla \omega _{L}/\gamma \protect \)
of external field; v) The mean distance between modes \protect\( \xi \nabla \omega _{L}\protect \)
in the units of frequency, \protect\( \xi \protect \) being the characteristic
spatial distance (\ref{lambda}) --- the wavelength of Airy type spin wave.
vi) \protect\( \omega _{M}=4\pi \gamma M\protect \) --- a characteristic of
the static demagnetizing field. vii) The radius \protect\( R\protect \), for
a box this is half-size, for factually used cylinders it is also roughly the
half-height. viii) -- x) The three parameters entering the Hamiltonian (\ref{f-ham})
used for simulations of the eigenvalue problem in a spherical cell, \protect\( (\xi /R)^{3}\protect \)
and \protect\( \omega _{M}/2R\nabla \omega _{L}\protect \) give the relative
importance correspondingly of the exchange molecular field and of the demagnetizing
field to the gradient of the external field; in a box these parameters lack
strict meaning and are presented here in parentheses only for an estimate; \protect\( C\protect \)
is the regime parameter (\ref{regimeC}) at the temperature shown for each experiment
in the left column. xi) The dipolar parameter (\ref{parameter}) (cylinder and
box) or (\ref{paramater-sph}) (sphere) (more exactly, their more accurate equivalents
(\ref{slab-corrections}) and (\ref{sphere-corrections}) respectively); if
it is much smaller than unity it signifies that perturbation theory can be applied
at the conditions of the experiment. The input parameters for the calculations
come from various sources \cite{table_explain}.}

\vspace{0.5cm}
\begin{tabular}{cccccccccccc}
\hline 
&
concen-&
magneti-&
Larmor&
 field&
 modes&
&
&
&
&
&
dipolar\\
&
tration&
zation&
frequency&
gradient&
distance&
&
&
&
&
&
 parameter\\
 parameters &
 \( x \),&
\( \ds {\frac{M}{M_{0}}} \)&
\( \ds {\frac{\omega _{L}}{2\pi }} \),&
\( \nabla H^{e} \),&
\( \ds {\frac{\xi \nabla \omega _{L}}{2\pi }} \),&
\( \ds {\frac{\omega _{M}}{2\pi }} \),&
\( R \),&
\( \ds {\frac{\xi }{R}} \),&
\( C \)&
\( \ds {\frac{\omega _{M}}{2R\nabla \omega _{L}}} \),&
(\ref{parameter}) or (\ref{paramater-sph})\\
 and units&
\( 10^{-3} \)&
&
 MHz&
 G/sm&
 Hz&
 Hz&
 mm&
\( \times 10^{-3} \)&
&
\( \times 10^{-3} \)&
\( \times 10^{-3} \)\\
\hline 
pure \( ^{3} \)He in a&
&
1&
1&
2&
 122&
 0.48&
&
(189)&
3.7&
(0.37)&
1.17\\
 rectangular box,&
-&
1&
 2&
 2&
97&
0.97&
1&
(150)&
7.3&
(0.74)&
2.5\\
 \cite{candela 1}, \( T=1 \) mK&
&
1&
 4&
 2&
77&
1.93&
&
(119)&
14.6&
(1.49)&
5.5\\
\\[-2ex] \( ^{3} \)He-\( ^{4} \)He in a sphere,&
 \( 1.82 \)&
1&
 258&
 3.62&
 60&
 9.2&
 0.4&
133&
2.5&
10.1&
38\\
 \cite{candela 2}, \( T=20 \) mK&
 \( 0.63 \)&
1&
258&
 2.01&
 30&
 3.9&
 0.6&
77&
3&
4.95&
28\\
 \\[-2ex]\( ^{3} \)He-\( ^{4} \)He in a &
&
1&
&
&
 78.4&
23.5&
&
121&
9.7&
18.1&
73\\
 hemisphere, \cite{Roni},&
&
2&
&
&
36.2&
47.0&
&
96&
19&
36.3&
173\\
 \( T=20 \) mK&
 93&
3&
 341&
 4&
 54.4&
70.6&
 0.5&
84&
29&
54.4&
288\\
&
&
4&
&
&
 49.4&
94.1&
&
76&
39&
72.5&
413\\
 \\[-2ex]pure \( ^{3} \)He in&
&
1&
&
&
 21.9&
 241&
&
6.6&
2.3&
36&
252\\
 a finite&
&
2&
&
&
 17.4&
 482&
&
5.2&
4.5&
73&
532\\
 cylinder, \cite{Roni},&
-&
3&
 312&
 5.1&
 15.2&
723&
 2.0&
4.6&
6.8&
109&
809\\
 \( T=20 \) mK&
&
4&
&
&
 13.8&
 963&
&
4.2&
9.1&
146&
1099\\
&
&
5&
&
&
 12.8&
1204&
&
3.9&
11&
182&
1385\\
 \\[-2ex]\( ^{3} \)He-\( ^{4} \)He in&
&
1&
&
&
150&
23.5&
&
22&
9.6&
1.7&
9.6\\
 a finite&
&
2&
&
&
119&
47.0&
&
17&
19&
3.4&
20\\
 cylinder, \cite{vermeulen},&
93&
3&
341&
10.6&
104&
70.6&
2.0&
15&
29&
5.1&
31\\
 \( T=20 \) mK&
&
4&
&
&
95&
94.6&
&
14&
39&
6.8&
42 \\
\hline 
\end{tabular}\end{table}

\end{widetext}

In particular, the influence of the demagnetizing field on the spectrum of standing
spin waves in an (\emph{infinite}) slab reduces for an arbitrary strength of
the dipolar field simply to uniform shift of all the modes by \( -\omega _{M}=-4\pi \gamma M \),
where \( M \) is the static magnetization of the liquid in an external field,
\( \gamma  \) --- gyromagnetic ratio.

In a \emph{finite} volume of a liquid strengthening of the demagnetizing field
results in the crossover from the Silin type spin wave regime to the regime
of magnetostatic waves. In this latter the mechanism of the forming of the standing
spin waves has nothing to do with the exchange physics of Fermi liquids. Magnetostatic
spectra are in no way specific to the Fermi liquid and so contain no information
on its parameters.

For intermediate polarizations the demagnetizing field in a finite volume of
a liquid leads to the corrections to Silin type spin wave spectra, both shifting
the spectra uniformly and also changing the distances between the modes and
the modes widths. These changes of the second type which distort the spectra
are undesirable from the point of view of deriving parameters of the liquid
from the spectra. 

We have seen that the parameter determining the influence of the demagnetizing
field on the spectra of spin waves in a finite cylinder in the first order of
perturbation theory is 
\begin{equation}
\label{parameter}
\frac{2\omega _{M}}{\pi \xi \nabla \omega _{L}}\frac{\xi }{R}\log \frac{\sqrt{R^{\Phi }L^{\Psi }}}{\xi ^{\Phi +\Psi }},
\end{equation}
 where \( \omega _{M}=4\pi \gamma M \) characterizes the magnetization density,
\( \nabla \omega _{L} \) is the gradient of the Larmor frequency, \( R \)
is the radius of the cylinder base, \( L \) its height and \( \xi  \) is the
wavelength (\ref{lambda}) of an Airy-type standing spin wave. The quantity
\( \xi \nabla \omega _{L} \) gives the average distance between modes in the
units of frequency. \( \Phi  \) and \( \Psi  \) are numbers of the order of
unity.

For a sphere, an analogous parameter was
\begin{equation}
\label{paramater-sph}
\frac{\sqrt{\pi }\omega _{M}}{4\xi \nabla \omega _{L}}\sqrt[4]{\frac{\xi }{2R}}.
\end{equation}
The values of these parameters (along with the values of some others having
appeared in the text) for several recent spin-wave experiments are arranged
into Table \ref{conditions}.

The ratio \( M/M_{0} \) of the absolute value \( M \) of the static magnetization
to the equilibrium magnetization \( M_{0} \) accounts for the possible higher
then equilibrium polarization of the liquid. Polarizing a liquid \( M/M_{0} \)
times its equilibrium value increases proportionally the strength of the demagnetizing
field as well as the parameters (\ref{parameter}), (\ref{paramater-sph}).
For the experiments \cite{vermeulen}, \cite{Roni}, where \( M/M_{0} \) could
be changed, we present in Table I the data for several integer values of \( M/M_{0} \)
which are close to the real experimental values.

For conditions when dipolar parameter (\ref{parameter}) or (\ref{paramater-sph})
is smaller than unity one can use perturbation theory to get corrections to
the modes frequencies. 

As the comparison between perturbation theory based analytical calculations
and numerical simulations for a sphere shows (see Fig. \ref{Fig-modesOnOmega})
perturbation theory has acceptable accuracy up to the values of \( \omega _{M}/2R\nabla \omega _{L}\sim 0.1 \)
or (\ref{parameter}), (\ref{paramater-sph}) \( \sim 0.4 \). However, at greater
values of the parameters (\ref{parameter}), (\ref{paramater-sph}) the discrepancy
between perturbation theory and numerics grows dramatically.

When the spin wave spectra are used for measuring the transverse relaxation
time \( \tau  \), a proper treatment of experimental data taking into account
dipolar field corrections is necessary. We estimated that the dipolar restrictions
on the correct determination of \( \tau  \) from the conventional interpretation
of the spectra are of the order of the parameter (\ref{parameter}) or (\ref{paramater-sph}).
In particular, for the experiment~\cite{vermeulen} about 4.2\%. The latter
means that the effect of a dipolar field can not significantly change the main
conclusion of this article that the polarization induced zero temperature spin
wave damping does not exist, which is in disagreement with previous spin echo
experiments \cite{Wei}, \cite{candela  OB}, \cite{Ager}, \cite{OB}.

A major inference for planning future experiments is the proposal to use ellipsoidal,
in particular, spherical containers, not only because the estimation of the
shape-dependent dipolar field effects is simpler, but also because there are
two roughly equal contributions to the spin wave spectrum distortion, --- one
from the inhomogeneity of the static demagnetizing field and the other from
the demagnetizing field produced by the rotating part of the magnetization.
The first contribution exists only in non-ellipsoidal samples, in which a homogeneous
static magnetization produces inhomogeneous demagnetizing field. For this reason
implementation of such shapes for the experiments on the elicitation of the
liquids characteristics from the spectra is disadvantageous.

The dependencies of (\ref{parameter}), (\ref{paramater-sph}) on experimentally
controllable parameters are as follows (\ref{parameter})\( \propto \omega _{L}\log \omega _{L} \),
\( \propto \log (D_{0}/\kappa \tau _{1}) \), \( \propto \nabla \omega _{L}^{-1}\log \nabla \omega _{L} \),
\( \propto R^{-1} \), and (\ref{paramater-sph})\( \propto \omega ^{5/4}_{L} \),
\( \propto (D_{0}/\kappa \tau _{1})^{-1/4} \), \( \propto \nabla \omega _{L}^{-3/4} \),
\( \propto R^{-1/4} \). So a bigger \( R \) and a bigger gradient diminish
the contribution of the demagnetizing field. 

But for typical experimental conditions as one can see from Table \ref{conditions},
strongly-polarized \( ^{3} \)He-\( ^{4} \)He solutions never are too far beyond
the regime of Silin spin waves perturbed by demagnetizing field, whereas pure
\( ^{3} \)He at strong polarizations is in the ``deep intermediate'' regime,
for which the results of Sec. V apply. In view of this, even at most favorable
cell size and field gradient, pure highly-polarized \( ^{3} \)He seems to be
unsuitable for a study of Silin waves spectra.

\section*{Acknowledgments}

The authors are grateful to Prof. I. A. Fomin for sharing with them his preliminary
view of the problem. One of us (P.~K.) also thanks the staff of the CRTBT for
their hospitality.

The work was financially supported by the Program ''Jumelage entre ENS et Institut
Landau'' and the Landau Scholarship from Forschungszentrum J\"{u}lich, Germany
(P.~K.).

\appendix

\section{Spin waves in an infinite medium in a uniform magnetic field\label{AppInfinite}}

The non-local dipolar term \( \underline{\widehat{n}}[\mathbf{m}] \) is known
to become local in the important case of a medium infinite in two directions,
with the proviso that \( \mathbf{m} \) depends on only the remaining third
coordinate. The direction in which \( \mathbf{m} \) varies we denote as \( \hat{s} \),
and then the medium should be infinite in the two directions perpendicular to
\( \hat{s} \).

In such conditions
\begin{equation}
\label{didj}
\partial _{i}\partial _{j}\int\limits _{V}\frac{f(\mathbf{r}')}{|\mathbf{r}-\mathbf{r}'|}d^{3}\mathbf{r}'=\hat{s}_{i}\hat{s}_{j}\partial _{s}\int \partial _{s}\hat{\mathcal{G}}_{1\mathrm{d}}(s-s')f(s')ds',
\end{equation}
 where by definition 
\begin{eqnarray*}
\partial _{a}\hat{\mathcal{G}}_{1\mathrm{d}}(a) & = & \int\limits ^{\infty }_{-\infty }\partial _{a}\frac{d^{2}\mathbf{r}_{\bot }'}{\sqrt{\mathbf{r}_{\bot }'^{2}+a^{2}}}=-\pi a\int\limits ^{\infty }_{a^{2}}\frac{dy}{y^{3/2}}=-2\pi \mathrm{sgn}a
\end{eqnarray*}
 Here \( \mathbf{r}_{\bot } \) denotes the coordinate vector in the plane perpendicular
to \( \hat{s} \).

So \( \underline{\widehat{n}}_{ij}[f]=\hat{s}_{i}\hat{s}_{j}f \) and
\begin{equation}
\label{n_{i}j}
\underline{\widehat{n}}[\mathbf{m}]=\hat{s}(\hat{s}\mathbf{m}).
\end{equation}

The case of an infinite medium in a uniform magnetic field (i.e. \( \nabla \omega _{L}=0 \))
is the simplest. Looking for a solution of (\ref{motion}) in the form of a
running wave 
\begin{equation}
\mathbf{m}(\mathbf{r})=\mathbf{m}_{0}e^{i\mathbf{kr}}
\end{equation}
 we have \( \hat{s}=\hat{k} \). Hence, Eq.~(\ref{motion}) from which the components
of the constant \( \mathbf{m}_{0} \) are to be found, becomes a linear algebraic
equation
\[
\left( \begin{array}{cc}
\widetilde{\omega }_{L}+\omega _{M}\hat{k}^{2}_{x} & -i\widetilde{\omega }+\omega _{M}\hat{k}_{x}\hat{k}_{y}\\
i\widetilde{\omega }+\omega _{M}\hat{k}_{x}\hat{k}_{y} & \widetilde{\omega }_{L}+\omega _{M}\hat{k}^{2}_{y}
\end{array}\right) \mathbf{m}_{0}=\frac{\omega _{M}}{4\pi }\mathbf{h}^{e}_{\bot },\]
 where \( \widetilde{\omega }_{L}=\omega _{L}-Dk^{2} \) and \( \widetilde{\omega }=\omega +i(D/C)k^{2} \).

\( \mathbf{m}_{0} \) as a function of frequency has resonance at the \( \omega  \)
which render the determinant of the matrix zero. This gives a Holstein-Primakoff
type spectrum with an additional attenuation term due to dissipation \( C^{-1}\neq 0 \)
\begin{equation}
\omega =\sqrt{\left( \omega _{L}-Dk^{2}\right) \left( \omega _{L}-Dk^{2}+\omega _{M}\sin ^{2}\theta \right) }-i\frac{D}{C}k^{2},
\end{equation}
 where \( \theta  \) is the angle between \( \widehat{z} \) and \( \mathbf{k} \).

\section{Spin wave spectrum\label{AppSpinSpectrum}}

A specimen placed into the field of an NMR coil changes its impedance in two
ways. First, the inductance \( L \) alters because so does the average energy
of the field \cite{LL}
\begin{equation}
\frac{LI^{2}}{2c^{2}}=\frac{1}{8\pi }\int \overline{\mathbf{H}(t,\mathbf{r})\mathbf{B}(t,\mathbf{r})}d^{3}\mathbf{r}
\end{equation}
 due to dispersion. \( I \) is the current through the coil. A line over an
expression here designates time average over oscillation period.

Secondly, the resistance \( R \) appears owing to the dissipation of the energy
of the field in the specimen 
\begin{equation}
RI^{2}=\frac{1}{4\pi }\int \overline{\mathbf{H}(t,\mathbf{r})\partial _{t}\mathbf{B}(t,\mathbf{r})}d^{3}\mathbf{r}.
\end{equation}

Making use of the general solution (\ref{B-dip-2})--(\ref{B-dip}) and of the
expansion (\ref{linear}) we write
\begin{eqnarray}
\mathbf{H}(t) & = & H^{e}\widehat{z}+\mathbf{h}^{e}(t)+\mathbf{H}_{\mathrm{dip}},\\
\mathbf{B}(t) & = & (H^{e}+4\pi M)\widehat{z}+\mathbf{h}^{e}(t)+4\pi \mathbf{m}(t)+\mathbf{H}_{\mathrm{dip}}.
\end{eqnarray}
 In taking space integrals of mutual scalar products of different terms of \( \mathbf{H}(t) \)
and \( \mathbf{B}(t) \) we note that those containing \( \mathbf{H}_{\mathrm{dip}} \)
transform into integrals over a remote surface. Since beyond the specimen magnetization
is zero, such integrals vanish.

In bilinear expressions we should write the monochromatic rf field \( \mathbf{h}^{e}(t)=\mathbf{h}_{\mathrm{m}}^{e}\cos \omega t \)
as \( \mathbf{h}^{e}(t)=\frac{1}{2}(\mathbf{h}^{e}+\mathbf{h}^{e*}) \), where
\( \mathbf{h}^{e}=\mathbf{h}_{\mathrm{m}}^{e}e^{-i\omega t} \). Similarly,
the rotating part of the magnetization should be written in the form \( \mathbf{m}(t)=\frac{1}{2}(\mathbf{m}+\mathbf{m}^{*}) \),
where \( \mathbf{m}=m_{\mathrm{m}}\mathrm{e}^{-i(\omega t+\varphi )}(1,i) \).

Then the impedance \( Z=R-i\omega L/c^{2} \) of an NMR coil can be written
in the form
\begin{eqnarray}
Z & = & Z_{0}+I^{-2}\int \overline{\mathbf{h}^{e}(t,\mathbf{r})\left( \partial _{t}-i\omega \right) \mathbf{m}(t,\mathbf{r})}d^{3}\mathbf{r}\nonumber \\
 & = & Z_{0}-i\omega I^{-2}\int \mathbf{h}^{e*}\mathbf{m}d^{3}\mathbf{r},
\end{eqnarray}
 where \( Z_{0} \) is the impedance without sample. So the change in the impedance
of the coil due to sample is proportional to 
\begin{equation}
2\int \mathbf{h}^{e*}\mathbf{m}d^{3}\mathbf{r}\approx \langle h^{e+}|m^{+}\rangle .
\end{equation}
 The real part of this quantity gives the dispersion spectrum, while imaginary
--- absorption.

Introducing Green operator \( \widehat{\mathcal{G}}_{\omega } \) (\ref{Green-operator})
and normalizing, we arrive at expression (\ref{self-inductance}).

\section{Dipolar corrections to modes in a finite cylinder\label{AppDipFinCyl}}

In this Appendix we derive expression (\ref{slab-corrections}) for mode shifts
due to dipolar field in a finite cylindrical cell. 

The normalization coefficients of transversely homogeneous \( n_{\rho }=m=0 \)
modes (\ref{trhmodes}) are 
\begin{equation}
c^{-2}_{n_{z}}=\pi R^{2}\xi \int ^{L/\xi \rightarrow \infty }_{0}\mathrm{Ai}^{2}\left( x+\alpha '_{n_{z}}\right) dx.
\end{equation}
 The upper limit may be put equal to infinity and then the last dimensionless
integral is a number depending only on \( n_{z} \).

In calculating the integrals in (\ref{ddip-cyl}) we use the expansion of the
Green function in cylindrical coordinates \cite[p. 140]{Jackson}
\begin{equation}
\label{r-r'expansion}
\frac{1}{|\mathbf{r}-\mathbf{r}'|}=\sum ^{+\infty }_{m=-\infty }e^{im(\varphi -\varphi ')}\int ^{\infty }_{0}e^{-k|z-z'|}J_{m}(k\rho )J_{m}(k\rho ')dk.
\end{equation}

Integrals over \( \varphi  \) and \( \varphi ' \) give \( (2\pi )^{2}\delta _{m0} \).
Then the integrals over \( \rho  \) and \( \rho ' \) with \( J_{0} \) give
\( R^{2}J^{2}_{1}(kR)/k^{2} \) so that 
\begin{eqnarray}
\int _{V} & \Psi  & (z)\underline{\widehat{n}}_{zz}\left[ \Phi (z)\right] d^{3}\mathbf{r}=\\
 & - & \pi R^{2}\int\limits ^{\infty }_{0}\frac{J^{2}_{1}(kR)}{k^{2}}\int\limits ^{L}_{0}\Psi (z)\int\limits ^{L}_{0}\Phi (z')\partial ^{2}_{z}e^{-k|z-z'|}dz'dzdk.\nonumber 
\end{eqnarray}
 In the first integral in (\ref{ddip-cyl}) \( \Psi (z)=\psi ^{2}_{n_{z}}(z) \),
\( \Phi (z)=1 \), while in the second --- \( \Psi (z)=\Phi (z)=\psi _{n_{z}}(z) \).

As a result of differentiating expansion (\ref{r-r'expansion}) we have 
\begin{equation}
\label{dxekx}
\partial ^{2}_{z}e^{-k|z-z'|}=-2k\delta (z-z')+k^{2}e^{-k|z-z'|}.
\end{equation}
 The integral of the \( \delta  \)-functional part is the simpler, using \( \int x^{-1}J^{2}_{1}(x)dx=\frac{1}{2} \),
we obtain \( \pi R^{2}\int ^{L}_{0}\Psi (z)\Phi (z)dz \). For both integrals
in (\ref{ddip-cyl}) this gives unity.

So the \( \delta  \)-functional part in the dipolar operator for transversely
homogeneous spatial distributions gives the local slab value (\ref{n-reduces-to-1}).

The second part in (\ref{dxekx}) is shown below to be non-zero only for finite
samples. Indeed, it yields 
\[
-\pi R\int ^{L}_{0}\Psi (z)\int ^{L}_{0}\Phi (z')F\left( \frac{|z-z'|}{R}\right) dz'dz,\]
 where 
\begin{equation}
F(p)=\int ^{\infty }_{0}J^{2}_{1}(x)e^{-px}dx
\end{equation}
 is the Laplace transform of \( J^{2}_{1} \). Although its value can be found
in tables (see, e.g. \cite[formula 6.612]{GR}) for our purposes it is sufficient
to know its value for small \( p \), where it diverges logarithmically 
\begin{equation}
F\left( \frac{|z-z'|}{R}\right) \approx -\frac{1}{\pi }\log \frac{e^{2}|z-z'|}{8R}.
\end{equation}

Writing for small \( z \) 
\[
\int ^{L}_{0}\log \frac{e^{2}|z-z'|}{8R}dz'=L\log \frac{eL}{8R}+z\log \frac{z}{eL}+O(z^{2})\]
 and substituting \( \frac{L-z}{\xi }\rightarrow x \) we obtain for the first
integral in (\ref{ddip-cyl}) 
\begin{equation}
\label{1st-order-2}
1+\frac{L}{\pi R}\log \frac{eL}{8R}+\frac{\Psi _{n_{z}}}{\pi }\frac{\xi }{R}\log \frac{\Theta _{n_{z}}\xi }{Le}.
\end{equation}

And for the second integral in (\ref{ddip-cyl}) 
\begin{equation}
\label{1st-order-1}
1+\frac{2\Phi _{n_{z}}}{\pi }\frac{\xi }{R}\log \frac{e^{2}\Xi _{n_{z}}\xi }{8R}.
\end{equation}

Here the \( n_{z} \)-dependent constants 
\begin{eqnarray}
\Phi _{n_{z}} & = & \frac{1}{2}\frac{(\int ^{\infty }_{-\alpha '_{n_{z}}}\mathrm{Ai}(x)dx)^{2}}{\int ^{\infty }_{-\alpha '_{n_{z}}}\mathrm{Ai}^{2}(x)dx},\\
\log \Xi _{n_{z}} & = & \frac{\int ^{\infty }_{-\alpha '_{n_{z}}}\int ^{\infty }_{-\alpha '_{n_{z}}}\log \left| x-x'\right| \mathrm{Ai}(x)\mathrm{Ai}(x')dxdx'}{(\int ^{\infty }_{-\alpha '_{n_{z}}}\mathrm{Ai}(x)dx)^{2}}\\
\Psi _{n_{z}} & = & \frac{\int ^{\infty }_{-\alpha '_{n_{z}}}x\mathrm{Ai}^{2}(x)dx}{\int ^{\infty }_{-\alpha '_{n_{z}}}\mathrm{Ai}^{2}(x)dx},\\
\log \Theta _{n_{z}} & = & \frac{\int ^{\infty }_{-\alpha '_{n_{z}}}x\log (x)\mathrm{Ai}^{2}(x)dx}{\int ^{\infty }_{-\alpha '_{n_{z}}}x\mathrm{Ai}^{2}(x)dx}
\end{eqnarray}
 are of the order unity, as is seen from Table \ref{TableFinCyl}, where they
are calculated numerically for the first six modes.

\begin{table}

\caption{Numerical constants in the expression (\ref{slab-corrections}) for mode shifts
due to dipolar field as functions of the mode number.\label{TableFinCyl}}
\begin{tabular}{p{0.9cm}p{1.1cm}p{1.1cm}p{1.1cm}p{1.1cm}p{1.1cm}p{1.1cm}}
\hline 
\( n_{z} \)&
0&
1&
2&
3&
4&
5\\
\hline 
\( \Phi _{n_{z}} \)&
1.1197&
0.9377&
0.6949&
0.6449&
0.5717&
0.5443\\
 \( \Xi _{n_{z}} \)&
0.5431 &
0.3525&
0.2441&
0.2384&
0.2019&
0.1989\\
 \( \Psi _{n_{z}} \)&
0.6792&
2.165&
3.2134&
4.109&
4.915&
5.659\\
 \( \Theta _{n_{z}} \)&
0.9189&
2.481&
3.657&
4.667&
5.576&
6.419 \\
\hline 
\end{tabular}\end{table}

Plugging (\ref{1st-order-2}) and (\ref{1st-order-1}) into (\ref{ddip-cyl})
we get for the dipolar corrections to the modes frequencies in a finite cylinder
expression (\ref{slab-corrections}).

\subsection{Dipolar error to transverse relaxation time}

We are now in a position to estimate the error in the determination of the transverse
relaxation time because of the dipolar field.

As an example, we consider the experiment \cite{vermeulen}, wherein \( \tau  \)
was obtained from the regime parameter \( C, \) which, in its turn, was determined
from the slope of the linear fit to the dependence of modes half-widths \( \Im \omega _{\alpha } \)
on their position \( \Re (\omega _{\alpha }-\omega _{L}-L\nabla \omega _{L}) \).
In the collisionless regime in a finite cylinder, as is seen from (\ref{spectrum-slab})
and (\ref{xi-substitute}), the two quantities are related through 
\begin{equation}
\label{1/3C}
\frac{\Im \omega _{\alpha }}{\Re (\omega _{\alpha }-\omega _{L}-L\nabla \omega _{L})}\approx \frac{1}{3C}.
\end{equation}
 We now find the dipolar correction to this value.

For finite \( C^{-1} \) the scale \( \xi  \) should be replaced with the complex
\( \xi (1+i/C)^{1/3} \). This means that the dipolar field changes also imaginary
parts of the eigenfrequencies and hence half-widths of the modes. In the collisionless
regime (\( C\gg 1 \)) the imaginary part of the correction (\ref{slab-corrections})
equals 
\begin{equation}
\label{slab-Im-corrections}
\delta _{\mathrm{dip}}\Im \omega ^{\mathrm{cylinder}}_{n_{z}}=\frac{\omega _{M}}{3C}\frac{\xi }{\pi R}\bigl (\Phi _{n_{z}}\log \frac{8R}{e^{3}\Xi _{n_{z}}\xi }+\Psi _{n_{z}}\log \frac{L}{\Theta _{n_{z}}\xi }\bigr ).
\end{equation}

So the dipolar field changes the widths of the modes proportionally to \( \omega _{M} \).
Since for all parameters being positive the initial imaginary parts \( \frac{1}{3C}\alpha _{n_{z}}'\xi \nabla \omega _{L} \)
of the modes (\ref{spectrum-slab}) are negative, the modes narrow down in the
first approximation.

As a matter of fact, experimentally measured are not the absolute positions
of the modes frequencies but rather their positions relative to each other.
So modes positions in Ref.~\cite{vermeulen} were determined relatively to the
position \( \omega _{0} \) of the zeroth mode: not \( \omega _{\alpha } \)
but \( \omega _{\alpha }-\omega _{0} \).

To diminish the error due to the data scattering in the value of the slope derived
from the fitting, it is desirable to fix the Larmor frequency at the wall \( \omega _{L}+L\nabla \omega _{L} \)
which is experimentally badly determinable. So the position \( \omega _{0} \)
was put equal to its value \( \omega ^{(0)}_{0}=\omega _{L}+L\nabla \omega _{L}+\alpha _{0}'\xi \nabla \omega _{L} \)
in the absence of the dipolar field. This means that instead of \( \omega _{\alpha }-\omega _{L}-L\nabla \omega _{L} \)
in the denominator of (\ref{1/3C}) actually used in Ref.~\cite{vermeulen}
was \( \Re ((\omega _{\alpha }-\omega _{0})+\alpha _{0}'\xi \nabla \omega _{L})\equiv \Re ((\omega _{\alpha }-\omega _{0})-\omega _{L}-L\nabla \omega _{L}+\omega ^{(0)}_{0}) \).

Calculating the ratio of \( \Im \omega _{\alpha } \) to this quantity taking
into account the dipolar corrections (\ref{slab-corrections}), (\ref{slab-Im-corrections})
we get 
\begin{eqnarray}
\frac{1}{3C}\Bigl (1 & - & \frac{\omega _{M}}{\alpha '_{n_{z}}\xi \nabla \omega _{L}}\frac{\xi }{\pi R}\bigl (\Phi _{0}\log \frac{8R}{e^{2+\Phi _{n_{z}}/\Phi _{0}}\Xi _{0}\xi }\nonumber \\
 & + & \Psi _{0}\log \frac{e^{1-\Psi _{n_{z}}/\Psi _{0}}L}{\Theta _{0}\xi }\bigr )\Bigr ).
\end{eqnarray}

So the error introduced to the determination of \( C \) is of the order of
(\ref{parameter}).

\section{Solutions for a spherical cell in adiabatic approximation\label{AppSph}}

In this Appendix we obtain spin wave solutions in a sphere neglecting the dipolar
field. 

For a solution of the three-dimensional eigenvalue problem in a sphere of a
radius \( R \)
\begin{eqnarray}
\left[ D\partial ^{2}+z\nabla \omega _{L}\right] u(z,\mathbf{r}_{\bot }) & = & \delta \omega \, u(z,\mathbf{r}_{\bot }),\label{AD-task} \\
\left. \partial _{r}u\right| _{r=R} & = & 0,\label{AD-taskbc} 
\end{eqnarray}
 where we denoted \( \delta \omega =\omega -\omega _{L} \) for brevity, the
adiabatic approximation consists in the substitution 
\begin{equation}
u(z,\mathbf{r}_{\bot })=v(z;\rho )w(\rho )e^{im\varphi },
\end{equation}
 where \( v_{n_{z}}(z;\rho ) \) are the eigenfunctions of the equation 
\begin{equation}
\label{AD-subtask1}
\left[ D\partial _{z}^{2}+z\nabla \omega _{L}\right] v_{n_{z}}(z;\rho )=\omega _{n_{z}}(\rho )v_{n_{z}}(z;\rho ).
\end{equation}

\begin{figure}
\resizebox*{0.8\columnwidth}{!}{\includegraphics{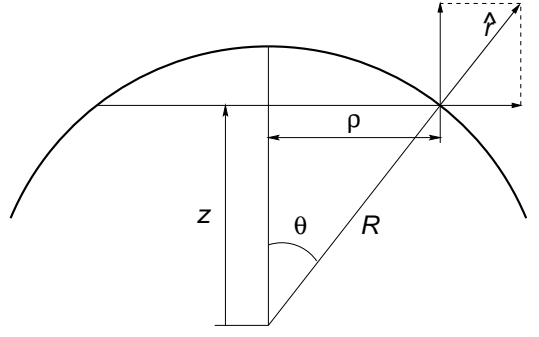}}

\caption{\label{scheme}A scheme of the coordinate system used to solve the eigenvalue
problem in the adiabatic approximation.}
\end{figure}

The coordinate notations are represented in Fig.~\ref{scheme}. The appropriate
boundary conditions for (\ref{AD-subtask1}) will be discussed below.

Then Eq.~(\ref{AD-task}) becomes 
\begin{eqnarray}
v_{n_{z}}(z;\rho )\left[ D\partial ^{2}_{\bot }+\omega _{n_{z}}(\rho )-\delta \omega \right] w(\rho )e^{im\varphi } &  & \nonumber \\
+De^{im\varphi }\left( w\partial ^{2}_{\bot }v_{n_{z}}+2\partial _{\rho }w\partial _{\rho }v_{n_{z}}\right)  & = & 0.\label{AD-neglect} 
\end{eqnarray}
 Adiabatic approximation utilizes the fact that the second term could be for
certain conditions neglected with respect to \( D\partial ^{2}_{\bot }w(\rho ) \).
Then we would obtain an effectively decoupled eigenvalue problem describing
the slow transverse motion
\begin{equation}
\label{AD-transverse}
\left[ D\partial ^{2}_{\bot }+\omega _{n_{z}}(\rho )\right] w(\rho )e^{im\varphi }=\delta \omega \, w(\rho )e^{im\varphi }.
\end{equation}

For an estimate of the conditions of the possibility of that neglect we write
\begin{equation}
\partial _{\rho }v_{n_{z}}(z;\rho )\sim \partial _{\rho }z\partial _{z}v_{n_{z}}\sim (\rho /R)\partial _{z}v_{n_{z}},
\end{equation}
 and from (\ref{AD-subtask1}) we see that \( \partial _{z}v_{n_{z}}\sim v_{n_{z}}/\xi  \),
where \( \xi  \) is the Airy spin wave characteristic wavelength (\ref{lambda}).
On the other hand \( \partial _{\rho }w\sim w/\rho  \). So, the condition of
applicability of the adiabatic approximation to this problem is that the ratio
of the second term in (\ref{AD-neglect}) to \( D\partial ^{2}_{\bot }w(\rho ) \)
should be smaller than unity
\begin{equation}
\rho ^{2}/R\xi \ll 1.
\end{equation}

The boundary condition (\ref{AD-taskbc}) on \( u \) reads 
\begin{equation}
[\cos \theta \partial _{z}+\sin \theta \partial _{\rho }]v_{n_{z}}(z;\rho )w(\rho )e^{im\varphi }=0.
\end{equation}
 For small \( \theta  \) we have \( \cos \theta \approx 1-\rho ^{2}/2R^{2} \)
and \( \sin \theta \approx \rho /R \). Estimating \( \partial _{\rho }v_{n_{z}} \)
as above, we see that \( \sin \theta \partial _{\rho }v_{n_{z}} \) is of the
order of \( (\rho /R)^{2}\partial _{z}v_{n_{z}} \), and in first order in \( \rho /R \)
the boundary conditions are 
\begin{equation}
\left( w\partial _{z}v_{n_{z}}+\frac{\rho }{R}v_{n_{z}}\partial _{\rho }w\right) _{z\approx R-\frac{\rho ^{2}}{2R}}=0.
\end{equation}

The first term is proportional to \( 1/\xi  \), while the second to \( 1/R \)
and in the case \( \xi \ll R \) the boundary conditions reduce to a much simpler
form 
\begin{equation}
\label{AD-bc}
\left. \partial _{z}v_{n_{z}}\right| _{z\approx R-\frac{\rho ^{2}}{2R}}=0
\end{equation}

Finally we formulate once again all our assumptions, i.e. \( \rho  \), \( \xi  \),
\( \rho ^{2}/\xi \ll R \). These three reduce to \( \xi \ll R \) if \( \rho \sim 1/\sqrt{a} \),
where \( 1/\sqrt{a} \) is the characteristic transverse spatial scale of the
wave function (see Eq.~(\ref{a}) below).

Now we proceed to the solution itself. First, Eq.~(\ref{AD-subtask1}) should
be solved with the boundary conditions (i) (\ref{AD-bc}) at \( z=R-\rho ^{2}/2R \)
and (ii) \( v_{i}\rightarrow 0 \) as \( z\rightarrow -\infty  \).

The solution is then 
\begin{equation}
\label{sphere-eigen}
v_{n_{z}}(z;\rho )=\mathrm{Ai}\left( \frac{\omega _{n_{z}}(\rho )-z\nabla \omega _{L}}{\xi \nabla \omega _{L}}\right) ,
\end{equation}
 where 
\begin{equation}
\omega _{n_{z}}(\rho )=\left( R-\frac{\rho ^{2}}{2R}+\xi \alpha _{n_{z}}'\right) \nabla \omega _{L}
\end{equation}
 and \( \alpha _{n}'<0 \) is the \( n \)-th zero of the derivative of the
Airy function \( \mathrm{Ai}' \).

Then Eq.~(\ref{AD-transverse}) describes a two-dimensional harmonic oscillator
\begin{eqnarray}
\left[ D\partial ^{2}_{\bot }-\frac{\nabla \omega _{L}}{R}\frac{\rho ^{2}}{2}\right] w(\rho )e^{im\varphi } & = & \nonumber \\
\left( \omega -\left( \alpha _{n_{z}}'\xi +R\right) \nabla \omega _{L}\right) w(\rho )e^{im\varphi }. &  & 
\end{eqnarray}

Solution in polar coordinates is 
\begin{equation}
\label{sphere-eigen-2}
w_{n_{\rho }m}(\rho )=\rho ^{|m|}e^{-a\rho ^{2}/2}L^{|m|}_{n_{\rho }}(a\rho ^{2}),
\end{equation}
 where \( L^{\alpha }_{n}(z) \) are the Laguerre polynomials and we denoted
\begin{equation}
\label{a}
a^{2}=\frac{\nabla \omega _{L}}{2DR}\equiv \frac{1}{2\xi ^{3}R}.
\end{equation}

The corresponding eigenfrequencies are 
\[
\omega _{n_{z}n}=\omega _{L}+R\nabla \omega _{L}+\xi \nabla \omega _{L}\left[ \alpha _{n_{z}}'-\sqrt{2\frac{\xi }{R}}(n+1)\right] ,\]
 where \( \omega _{L} \) is the Larmor frequency in the center of the sphere,
\( n=2n_{\rho }+|m| \). 

Again we remark that only the modes with zero azimuthal quantum number \( m \)
couple to the rf field. Eigenfunctions with \( m=0 \) occur for even \( n=2n_{\rho } \),
as is written in (\ref{spectrum-sphere}).

\subsection{Effects of dipolar field in the first order of perturbation theory}

We now calculate the demagnetizing field corrections (\ref{ddip-cyl}) to the
modes in a sphere.

As we saw in the previous subsection, the dipolar-free solution of the equations
of motion satisfying boundary conditions (\ref{boundary-condition}) is 
\begin{equation}
\psi _{n_{z}n_{\rho }0}=c_{n_{z}n_{\rho }0}\mathrm{Ai}\bigl (\ts {\frac{R-\frac{\rho ^{2}}{2R}-z}{\xi }}+\alpha '_{n_{z}}\bigr )e^{-a\rho ^{2}/2}L_{n_{\rho }}(a\rho ^{2}),
\end{equation}
 where \( n_{z},n_{\rho }=0,1,2,...,\infty  \), and \( m \) was put equal
to zero because only the modes with \( m=0 \) couple to the homogeneous rf
field. Furthermore, for simplicity we consider modes with \( n_{\rho }=0 \)
for which \( L_{n_{\rho }}=1 \).

In order to calculate the normalization coefficient \( c_{n_{z}00} \)
\[
c^{-2}_{n_{z}00}=2\pi \int ^{\infty }_{0}e^{-a\rho ^{2}}\int ^{R-\frac{\rho ^{2}}{2R}}_{-R+\frac{\rho ^{2}}{2R}}\mathrm{Ai}^{2}\bigl (\ts {\frac{R-\frac{\rho ^{2}}{2R}-z}{\xi }}\bigr )dz\, \rho d\rho \]
 we make the substitution \( x=(R-\rho ^{2}/2R-z)/\xi  \)
\begin{eqnarray}
c^{-2}_{n_{z}00} & = & 2\pi \xi \int\limits ^{\infty }_{0}e^{-a\rho ^{2}}\! \! \! \! \! \! \! \! \int\limits ^{2\frac{R-\frac{\rho ^{2}}{2R}}{\xi }\rightarrow \infty }_{0}\! \! \! \! \! \! \! \! \mathrm{Ai}^{2}\left( x+\alpha '_{n_{z}}\right) dx\, \rho d\rho \nonumber \\
 & = & \frac{\pi \xi }{a}\int ^{\infty }_{0}\mathrm{Ai}^{2}\left( x+\alpha '_{n_{z}}\right) dx,
\end{eqnarray}
 Here the upper limit of integration over \( x \) may be estimated as \( \sim (R/\xi -1/aR\xi )=(R/\xi -\sqrt{2\xi /R})\gg 1 \),
after which the integral over \( x \) decouples from that over \( \rho  \)
in the approximation \( \xi /R\ll 1 \).

In the case of a sphere \( \underline{\widehat{n}}_{zz}[1]=n^{(z)}=\frac{1}{3} \)
and as was already mentioned there is no contribution to the spectrum distortion
from the static dipolar field. One needs only to calculate the second integral
in (\ref{ddip-cyl}).

Again, the \( \delta - \)functional term in (\ref{dxekx}) using \cite{Jackson}
\[
\int ^{\infty }_{0}kJ_{\nu }(k\rho )J_{\nu }(k\rho ')dk=\frac{1}{\rho }\delta (\rho -\rho ')\]
 can be easily seen to give a constant \( 1 \) as it should. In the limit \( \xi /R\rightarrow 0 \)
a sphere transforms into a half-space, the modes being localized near the boundary.
A half-space is a particular case of a slab, with the height \( L\rightarrow \infty  \).
Hence the solutions in a sphere in the limit \( \xi /R\rightarrow 0 \) transform
into solutions for a thick slab depending on only \( z \). The demagnetizing
tensor for such solutions reduces to a constant (\ref{n-reduces-to-1}).

The second term in (\ref{dxekx}) gives 
\begin{eqnarray}
\int\limits _{V} & \! \! \! \! \psi _{n_{z}} & \! \! \! \! (z)\underline{\widehat{n}}_{zz}\left[ \psi _{n_{z}}(z)\right] d^{3}\mathbf{r}=-\pi c^{2}_{n_{z}}\xi ^{2}\! \! \int\limits ^{\infty }_{0}\! \! k^{2}\int\limits ^{\infty }_{0}\! \! e^{-a\rho ^{2}/2}J_{0}(k\rho )\nonumber \\
 & \times  & \int\limits ^{\infty }_{0}\! \! e^{-a\rho '^{2}/2}J_{0}(k\rho ')\! \! \int\limits ^{\infty }_{0}\! \! \mathrm{Ai}(x+\alpha '_{n_{z}})\! \! \int\limits ^{\infty }_{0}\! \! \mathrm{Ai}(x'+\alpha '_{n_{z}})\nonumber \\
 & \times  & e^{-k\xi \left| x-x'+\frac{\rho ^{2}-\rho '^{2}}{2R\xi }\right| }dx'\, dx\, \rho 'd\rho '\, \rho d\rho \, dk.
\end{eqnarray}
 The two terms under the module sign appeared from \( |z-z'| \). We can estimate
\( x-x'\sim 1 \) and \( (\rho ^{2}-\rho '^{2})/2R\xi \sim 1/2R\xi a=\sqrt{\xi /2R} \),
therefore \( |z-z'|\approx \xi |x-x'| \). Then the integrals over \( \rho  \)
and \( \rho ' \) can be taken \cite[formula 6.631]{GR}
\begin{equation}
\int ^{\infty }_{0}e^{-a\rho ^{2}/2}J_{0}(k\rho )\rho d\rho =\frac{1}{a}e^{-k^{2}/2a}.
\end{equation}

The integral over \( k \)
\begin{equation}
\label{over-k}
\int ^{\infty }_{0}e^{-k^{2}/a}e^{-kp}k^{2}dk,
\end{equation}
 where \( p=\xi |x-x'| \), is a function of \( p\sqrt{a}\sim \sqrt[4]{\xi /R} \).
Therefore, we can substitute (\ref{over-k}) with the zeroth term of its expansion
in series with respect to \( p\sqrt{a} \), which is \( \frac{1}{4}\sqrt{\pi a^{3}} \).
Then
\begin{eqnarray}
\int _{V} & \psi _{n_{z}} & (z)\underline{\widehat{n}}_{zz}\left[ \psi _{n_{z}}(z)\right] d^{3}\mathbf{r}=1-\frac{\Phi _{n_{z}}}{2}\sqrt{\pi a}\xi \nonumber \\
 & = & 1-\frac{\Phi _{n_{z}}\sqrt{\pi }}{2}\sqrt[4]{\frac{\xi }{2R}},
\end{eqnarray}
 where \( \Phi _{n_{z}} \) are the same numbers as in the case of a cylinder.

In the end we obtain (\ref{sphere-corrections}) for the corrections to the
modes.

\section{Matrix elements of the demagnetizing operator\label{AppMatrixElements}}

In this Appendix we are going into the detail of calculation of the matrix elements
of the dipolar integro-differential operator \( \underline{\widehat{n}}_{zz} \).
We show that only those elements \( \langle nlm|\underline{\widehat{n}}_{zz}|n'l'm'\rangle  \)
are non-zero that are between the states with \( m=m' \) and \( l'=l,l\pm 2 \).

Before proceeding we remark on notations. We will write the integral operator
in (\ref{demag-tensor}) as
\begin{equation}
\hat{\mathcal{G}}_{\infty }\mathbf{M}(\mathbf{r})=\int \frac{\mathbf{M}(\mathbf{r}')}{|\mathbf{r}-\mathbf{r}'|}d^{3}\mathbf{r}',
\end{equation}
 justification being that \( \hat{\mathcal{G}}_{\infty } \) is the Green operator
for Laplace equation with the boundary condition of vanishing at infinity.

So our plan for this section is, first, to calculate \( \hat{\mathcal{G}}_{\infty }|nlm\rangle  \).
Then, secondly, we calculate the result of acting of \( \partial _{z}^{2} \)
on an arbitrary function \( f(\mathbf{r}) \) expanded in spherical harmonics
\begin{equation}
\label{f}
f(\mathbf{r})=\sum _{lm}f_{lm}(r)Y^{m}_{l}(\hat{\mathbf{r}})
\end{equation}

By substituting \( \hat{\mathcal{G}}_{\infty }|nlm\rangle  \) for \( f \)
we eventually find the matrix elements themselves.

\subsection{Calculation of \protect\( \hat{\mathcal{G}}_{\infty }|nlm\rangle \protect \)}

We find \( \hat{\mathcal{G}}_{\infty }|nlm\rangle  \) directly by integration,
using a well-known formula from the theory of spherical harmonics 
\[
\frac{1}{|\mathbf{r}-\mathbf{r}'|}=\sum _{lm}\frac{4\pi }{2l+1}Y^{m}_{l}(\hat{\mathbf{r}})Y^{m*}_{l}(\hat{\mathbf{r}}')\frac{r^{l}_{<}}{r^{l+1}_{>}},\]
 where \( r_{<} \) (\( r_{>} \)) is the smaller (larger) of \( r \) and \( r' \).
Then 
\begin{eqnarray}
\hat{\mathcal{G}}_{\infty }|nlm\rangle  & = & \frac{4\pi }{2l+1}c_{nl}Y^{m}_{l}(\hat{\mathbf{r}})\label{App-2} \\
 & \times  & \left( \int ^{r}_{0}\frac{r'^{l}}{r^{l+1}}+\int ^{R}_{r}\frac{r^{l}}{r'^{l+1}}\right) j_{l}(k_{nl}r')r'^{2}dr'.\nonumber 
\end{eqnarray}

The integrals at the right-hand side can be taken easily as follows. We notice
that \cite{abramovitz}
\begin{eqnarray}
x^{-l}j_{l+1}(x) & = & -\partial _{x}\left[ x^{-l}j_{l}(x)\right] ,\\
x^{l+1}j_{l-1}(x) & = & \partial _{x}\left[ x^{l+1}j_{l}(x)\right] .\label{j-l-1} 
\end{eqnarray}
 Taking integrals from the two sides we obtain

\begin{eqnarray}
\int ^{x}_{0}\left( \frac{y}{x}\right) ^{l+1}j_{l-1}(y)dy & = & j_{l}(x),\\
\int ^{a}_{x}\left( \frac{x}{y}\right) ^{l}j_{l+1}(y)dy & = & j_{l}(x)-\left( \frac{x}{a}\right) ^{l}j_{l}(a).
\end{eqnarray}

On plugging the above into (\ref{App-2}) and using another property of the
spherical Bessel functions
\begin{equation}
\label{j-l-1-l+1}
j_{l+1}\left( x\right) +j_{l-1}\left( x\right) =\frac{2l+1}{x}j_{l}(x)
\end{equation}
 we get a sum of two terms 
\begin{eqnarray}
\hat{\mathcal{G}}_{\infty }|nlm\rangle  & = & \frac{4\pi R^{2}}{(k_{nl}R)^{2}}|nlm\rangle \label{G-inf-nlm} \\
 & - & \frac{4\pi R^{2}}{2l+1}\frac{c_{nl}}{k_{nl}R}Y^{m}_{l}(\hat{\mathbf{r}})\left( \frac{r}{R}\right) ^{l}j_{l-1}(k_{nl}R).\nonumber \label{G-inf-nkm} 
\end{eqnarray}
 This expression is inapplicable when \( k_{nl}R=0 \), which takes place for
\( n=l=0 \). In this particular case, integrating explicitly, we get
\[
\hat{\mathcal{G}}_{\infty }|000\rangle =4\pi c_{00}Y^{0}_{0}(\hat{\mathbf{r}})j_{0}(0)\left( \frac{R^{2}}{2}-\frac{r^{2}}{6}\right) .\]

\subsection{Calculation of \protect\( \partial ^{2}_{z}f(\mathbf{r})\protect \) \label{transverse}}

Making use of the overt expressions of the basis vectors in the spherical coordinates
\begin{eqnarray}
\widehat{\mathbf{r}} & = & \widehat{\mathbf{x}}\sin \theta \cos \varphi +\widehat{\mathbf{y}}\sin \theta \sin \varphi +\widehat{\mathbf{z}}\cos \theta ,\nonumber \\
\widehat{\f \theta } & = & \widehat{\mathbf{x}}\cos \theta \cos \varphi +\widehat{\mathbf{y}}\cos \theta \sin \varphi -\widehat{\mathbf{z}}\sin \theta ,\\
\widehat{\f \varphi } & = & -\widehat{\mathbf{x}}\sin \varphi +\widehat{\mathbf{y}}\cos \varphi ,\nonumber 
\end{eqnarray}
 and of the nabla operator 
\begin{equation}
\f \partial =\widehat{\mathbf{r}}\partial _{r}+\widehat{\f \theta }\frac{1}{r}\partial _{\theta }+\widehat{\f \varphi }\frac{1}{r\sin \theta }\partial _{\varphi },
\end{equation}
 it can be verified that 
\begin{equation}
\label{dz}
\partial _{z}=\cos \theta \partial _{r}+\frac{1}{2r}\sin \theta \left( e^{i\varphi }\hat{l}_{-}-e^{-i\varphi }\hat{l}_{+}\right) .
\end{equation}

Here \( \hat{\mathbf{l}}=-i\mathbf{r}\times \f \partial  \) is the angular
momentum operator 
\begin{equation}
\hat{l}_{z}=-i\partial _{\varphi },\quad \hat{l}_{\pm }=e^{\pm i\varphi }(\pm \partial _{\theta }+i\cot \theta \partial _{\varphi }),
\end{equation}
 which has the well known effect on the spherical harmonics 
\begin{eqnarray}
\hat{l}_{z}Y_{l}^{m} & = & mY_{l}^{m},\\
\hat{l}_{+}Y_{l}^{m} & = & \sqrt{(l-m)(l+m+1)}Y_{l}^{m+1},\\
\hat{l}_{-}Y_{l}^{m} & = & \sqrt{(l+m)(l-m+1)}Y_{l}^{m-1}.
\end{eqnarray}

The product of each spherical harmonic with \( \sin \theta e^{i\varphi }\propto Y^{1}_{1} \),
\( \cos \theta \propto Y^{0}_{1} \), or \( \sin \theta e^{-i\varphi }\propto Y^{-1}_{1} \),
is a sum \cite{merzbacher}

\begin{eqnarray}
\textstyle \sin \theta e^{-i\varphi }Y^{m}_{l} & = & a^{m}_{l}Y^{m-1}_{l+1}-b^{m}_{l}Y^{m-1}_{l-1},\\
\sin \theta e^{i\varphi }Y^{m}_{l} & = & -a^{-m}_{l}Y^{m+1}_{l+1}+b^{-m}_{l}Y^{m+1}_{l-1},\\
\cos \theta Y^{m}_{l} & = & c^{m}_{l+1}Y^{m}_{l+1}+c^{m}_{l}Y^{m}_{l-1}\label{cos} 
\end{eqnarray}
 of the harmonics with the adjacent \( l \) and \( m \) multiplied each by
a coefficient (which, in fact, are particular cases of the Clebsh-Gordan coefficients)
\begin{eqnarray}
a^{m}_{l} & = & \sqrt{\ts {\frac{l-m+1}{2l+1}\frac{l-m+2}{2l+3}}},\nonumber \\
b^{m}_{l} & = & \sqrt{\ts {\frac{l+m}{2l+1}\frac{l+m-1}{2l-1}}},\label{ab} \\
c^{m}_{l} & = & \sqrt{\ts {\frac{l+m}{2l+1}\frac{l-m}{2l-1}}}.\nonumber 
\end{eqnarray}

Hence from (\ref{dz})
\begin{equation}
\label{d-f}
\partial _{z}f=\sum _{lm}\left( -c^{m}_{l+1}Y^{m}_{l+1}\hat{\mathcal{L}}_{l}^{+}+c^{m}_{l}Y^{m}_{l-1}\hat{\mathcal{L}}_{l}^{-}\right) f_{lm}(r).
\end{equation}
 Here the coefficients turn out to be the same as in (\ref{cos}), and we introduced
two differentiating operators 
\begin{eqnarray}
\hat{\mathcal{L}}_{l}^{+} & = & -\partial _{r}+l/r,\nonumber \\
\hat{\mathcal{L}}_{l}^{-} & = & \partial _{r}+(l+1)/r.\label{L} 
\end{eqnarray}

We may rewrite (\ref{d-f}) by shifting the summation indices as 
\[
\partial _{z}f(\mathbf{r})=\sum _{lm}(\partial _{z}f)_{lm}(r)Y_{l}^{m}(\hat{\mathbf{r}}),\]
 where
\begin{equation}
(\partial _{z}f)_{lm}=-c^{m}_{l}\hat{\mathcal{L}}_{l-1}^{+}f_{l-1,m}+c^{m}_{l+1}\hat{\mathcal{L}}_{l+1}^{-}f_{l+1,m}.
\end{equation}

Then repeating the procedure, we get 
\[
\partial ^{2}_{z}f(\mathbf{r})=\sum _{lm}(\partial ^{2}_{z}f)_{lm}(r)Y_{l}^{m}(\hat{\mathbf{r}}),\]
 with
\begin{eqnarray}
\left( \partial ^{2}_{z}f\right) _{lm} & = & c^{m}_{l}c^{m}_{l-1}\hat{\mathcal{L}}_{l-1}^{+}\hat{\mathcal{L}}_{l-2}^{+}f_{l-2,m}\\
 & + & \left[ \left( c^{m}_{l}\right) ^{2}+\left( c^{m}_{l+1}\right) ^{2}\right] \left( \partial ^{2}\right) _{l}f_{lm}\nonumber \\
 & + & c^{m}_{l+1}c^{m}_{l+2}\hat{\mathcal{L}}_{l+1}^{-}\hat{\mathcal{L}}_{l+2}^{-}f_{l+2,m}.\nonumber 
\end{eqnarray}
 Here we used that \( \hat{\mathcal{L}}_{l-1}^{+}\hat{\mathcal{L}}_{l}^{-}=\hat{\mathcal{L}}_{l+1}^{-}\hat{\mathcal{L}}_{l}^{+}=-\left( \partial ^{2}\right) _{l} \),
where \( \left( \partial ^{2}\right) _{l} \) is the \( l \)-th component of
the Laplace operator 
\begin{equation}
\left( \partial ^{2}\right) _{l}=\partial ^{2}_{r}+2\frac{\partial _{r}}{r}-\frac{l(l+1)}{r^{2}}.
\end{equation}

Changing the summation indices in each term again we arrive at
\begin{eqnarray}
\partial ^{2}_{z}f & =\sum _{lm} & \left( c^{m}_{l+1}c^{m}_{l+2}Y^{m}_{l+2}(\hat{\mathbf{r}})\hat{\mathcal{L}}_{l+1}^{+}\hat{\mathcal{L}}_{l}^{+}\right. \label{dzz} \\
 & + & \left[ \left( c^{m}_{l}\right) ^{2}+\left( c^{m}_{l+1}\right) ^{2}\right] Y^{m}_{l}(\hat{\mathbf{r}})\left( \partial ^{2}\right) _{l}\nonumber \\
 & + & \left. c^{m}_{l-1}c^{m}_{l}Y^{m}_{l-2}(\hat{\mathbf{r}})\hat{\mathcal{L}}_{l-1}^{-}\hat{\mathcal{L}}_{l}^{-}\right) f_{lm}\nonumber 
\end{eqnarray}

\subsection{Matrix elements}

\begin{table}

\caption{Operators \protect\( \hat{\mathcal{L}}_{l}^{+}\protect \), \protect\( \hat{\mathcal{L}}_{l}^{-}\protect \).\label{TableOperators}}
\begin{tabular}{p{1cm}p{1.7cm}p{1.7cm}p{1.2cm}p{2.5cm}}
\hline 
\( f(r) \)&
 \( r^{l} \)&
 \( 1/r^{l+1} \)&
 \( j_{l}(r) \)&
 \( j_{l}(\mathrm{const}\: r) \)\\
\hline 
\( \hat{\mathcal{L}}_{l}^{+}f \)&
 \( 0 \)&
 \( (2l+1)f/r \)&
 \( j_{l+1}(r) \)&
 \( \mathrm{const}\: j_{l+1}(\mathrm{const}\: r) \)\\
 \( \hat{\mathcal{L}}_{l}^{-}f \)&
 \( (2l+1)f/r \)&
 \( 0 \)&
 \( j_{l-1}(r) \)&
 \( \mathrm{const}\: j_{l-1}(\mathrm{const}\: r) \) \\
\hline 
\end{tabular}

\end{table}

Substituting \( \hat{\mathcal{G}}_{\infty }|nlm\rangle  \) from (\ref{G-inf-nlm})
into (\ref{dzz}) and using the properties of \( \hat{\mathcal{L}}_{l}^{+} \),
\( \hat{\mathcal{L}}_{l}^{-} \) summarized in Table \ref{TableOperators},
we find 
\begin{eqnarray}
\frac{1}{4\pi }\partial ^{2}_{z}\hat{\mathcal{G}}_{\infty }|nlm\rangle  & = & c^{m}_{l+1}c^{m}_{l+2}c_{nl}j_{l+2}(k_{nl}r)Y^{m}_{l+2}(\hat{\mathbf{r}})\nonumber \\
 & - & \left[ \left( c^{m}_{l}\right) ^{2}+\left( c^{m}_{l+1}\right) ^{2}\right] |nlm\rangle \\
 & + & c^{m}_{l-1}c^{m}_{l}c_{nl}Y^{m}_{l-2}(\hat{\mathbf{r}})\Biggl (j_{l-2}(k_{nl}r)\nonumber \\
 & - & (2l-1)\frac{j_{l-1}(k_{nl}R)}{k_{nl}R}\left( \frac{r}{R}\right) ^{l-2}\Biggl ).\nonumber 
\end{eqnarray}
 It is not hard to verify by straightforward differentiating that this expression
holds also for \( n=l=m=0 \).

The matrix elements \( \langle n'l'm'|\partial ^{2}_{z}\hat{\mathcal{G}}_{\infty }|nlm\rangle  \)
are to be obtained from this expression by integrating with the complex conjugate
of (\ref{mu}) over the spherical volume of our sample. In doing so we see that
only the elements with \( m'=m \), \( l'=l,l\pm 2 \) are non-zero as we already
mentioned in the text 
\begin{eqnarray}
\langle nlm| & \frac{1}{4\pi } & \partial ^{2}_{z}\hat{\mathcal{G}}_{\infty }|n',l'=l-2,m\rangle \nonumber \\
 & = & (c^{m}_{l-1}c^{m}_{l})\int ^{R}_{0}cc'j_{l}(kr)j_{l}(k'r)r^{2}dr;\nonumber \\
\langle nlm| & \frac{1}{4\pi } & \partial ^{2}_{z}\hat{\mathcal{G}}_{\infty }|n'lm\rangle =-\left[ \left( c^{m}_{l}\right) ^{2}+\left( c^{m}_{l+1}\right) ^{2}\right] \delta _{nn'};\nonumber \\
\langle nlm| & \frac{1}{4\pi } & \partial ^{2}_{z}\hat{\mathcal{G}}_{\infty }|n',l'=l+2,m\rangle \label{App-me} \\
 & = & (c^{m}_{l'-1}c^{m}_{l'})cc'\Biggl (\int ^{R}_{0}j_{l}(kr)j_{l}(k'r)r^{2}dr\nonumber \\
 &  & -(2l+3)R^{3}\frac{j_{l+1}(kR)}{kR}\frac{j_{l+1}(k'R)}{k'R}\Biggl ),\nonumber 
\end{eqnarray}
 where we introduced shorthand notations \( k=k_{nl} \), \( k'=k_{n'l'} \),
\( c=c_{nl} \), \( c'=c_{n'l'} \).

In the last term we used 
\begin{equation}
\label{App-int}
\int ^{R}_{0}j_{l}(kr)r^{l+2}dr/R^{l}=R^{3}j_{l+1}(kR)/kR
\end{equation}
 with \( k=k_{nl} \). There is one exception, though, when \( l=n=0 \) ---
then \( k_{nl}=k_{00}=0 \) and the division in the right-hand side of (\ref{App-int})
is undefined. Integrating straightforwardly, we get instead \( R^{3}j_{0}(0)/3 \).

At the end, we take integrals in the off-diagonal elements of (\ref{App-me})
using the formulas below \cite[formula 5.54]{GR}

\begin{eqnarray}
\int  & j_{l}( & ax)j_{l}(bx)x^{2}dx\label{ja-jb} \\
 & = & x^{2}\frac{bj_{l-1}(bx)j_{l}(ax)-aj_{l-1}(ax)j_{l}(bx)}{a^{2}-b^{2}}\nonumber \\
 & \equiv  & x^{2}\frac{-bj_{l+1}(bx)j_{l}(ax)+aj_{l+1}(ax)j_{l}(bx)}{a^{2}-b^{2}}\nonumber \\
 & \equiv  & x^{2}\frac{bj_{l+1}(bx)j_{l+2}(ax)-aj_{l+1}(ax)j_{l+2}(bx)}{a^{2}-b^{2}}\nonumber \\
 &  & +(2l+3)x^{3}\frac{j_{l+1}(ax)}{ax}\frac{j_{l+1}(bx)}{bx}.\nonumber 
\end{eqnarray}
 Here \( a\neq b \) and we consecutively applied recurrent relation (\ref{j-l-1-l+1})
first to \( j_{l-1} \) and then to \( j_{l} \).

Plugging the integration limits we see that above-diagonal (\( l'=l+2 \)) elements
equal corresponding below-diagonal (\( l'=l-2 \)) with the appropriate change
\( l'\leftrightarrow l \). (For \( n=l=0 \) this is to be verified manually.)

Finally, we get formula (\ref{me-dip}) in the text for the dipolar part of
the Hamiltonian.

The case \( a=b \) in (\ref{ja-jb}) is used in the calculation of the normalization
coefficients \( c_{nl} \) (\ref{sph-norm}) 
\[
\int j^{2}_{l}(ax)x^{2}dx=\frac{x^{3}}{2}\left( j^{2}_{l}(ax)-j_{l-1}(ax)j_{l+1}(ax)\right) .\]
 We get 
\begin{eqnarray*}
c_{nl} & = & \left( \frac{R^{3}}{2}\left[ j^{2}_{l}(k_{nl}R)-j_{l-1}(k_{nl}R)j_{l+1}(k_{nl}R)\right] \right) ^{-1/2},\\
 & = & \Bigl (\frac{R^{3}}{2}\bigl [j^{2}_{l}(k_{nl}R)+j^{2}_{l+1}(k_{nl}R)\\
 & - & \frac{2l+1}{k_{nl}R}j_{l}(k_{nl}R)j_{l+1}(k_{nl}R)\bigr ]\Bigr )^{-1/2},
\end{eqnarray*}
 when \( n+l\geq 1 \). For \( n=l=0 \) we have \( k_{00}=0 \) and \( j_{l-1} \)
grows unlimitedly when its argument tends to zero. Then we integrate directly
\( c_{00}=\left( R^{3}j^{2}_{0}/3\right) ^{-1/2}. \)


\begin{thebibliography}{10}
\bibitem{Meyerovich}A. E. Meyerovich, Phys. Lett. \textbf{A107}, 177 (1985). 
\bibitem{JeonMullin}J. W. Jeon and W. J. Mullin, J. Phys. (Paris) \textbf{49}, 1691 (1988); Phys.
Rev. Lett. \textbf{62}, 2691 (1989); W. J. Mullin and J. W. Jeon, J. Low Temp.
Phys. \textbf{88}, 433 (1992). 
\bibitem{MeyerovichMusaelian}A. E. Meyerovich and K. A. Musaelian, J. Low Temp. Phys. \textbf{89}, 781 (1992);
J. Low Temp. Phys. \textbf{94}, 249 (1994); J. Low Temp. Phys. \textbf{95},
789 (1994); Phys. Rev. Lett. \textbf{72}, 1710 (1994). 
\bibitem{Golosov}D. I. Golosov and A. E. Ruckenstein, Phys. Rev. Lett., \textbf{74}, 1613 (1995);
J. Low Temp. Phys. \textbf{112}, 265 (1998). 
\bibitem{Fomin-solo}I. A. Fomin, JETP Lett., \textbf{65}, 749 (1997). 
\bibitem{Wei}D. Candela, D. R. McAllaster, L-J. Wei, and N. Kalechofsky, J. Low Temp. Phys.
\textbf{89}, 307 (1992); L.-J. Wei, N. Kalechofsky and D. Candela, Phys. Rev.
Lett. \textbf{71}, 879 (1993). 
\bibitem{candela  OB}D. Candela, H. Akimoto, R. M. Bowley, O. Buu, D. Clubb, J. R. Owers-Bradley,
J. Low Temp. Phys. \textbf{121}, 767 (2000).
\bibitem{Ager}J. H. Ager, A. Child, R. Konig et al., J. Low Temp. Phys. \textbf{99}, 683 (1995). 
\bibitem{OB}J. R. Owers-Bradley, R. M. Bowley, O. Buu, D. Clubb and G. Vermeulen, J. Low
Temp. Phys. \textbf{121}, 779 (2000).
\bibitem{vermeulen}G. A. Vermeulen, A. Roni, Phys. Rev. Lett. \textbf{86}, 248 (2001). 
\bibitem{golubev98}D. S. Golubev and A. D. Zaikin, Phys. Rev. Lett. \textbf{81}, 1074 (1998), Phys.
Rev. B \textbf{59}, 9195 (1999).
\bibitem{aleiner99}I. L. Aleiner, B. L. Altshuler, M. E. Gershenson, Phys. Rev. Lett. \textbf{82},
3190 (1999). 
\bibitem{cohen99}D. Cohen, J. Imry, Phys. Rev. B \textbf{59}, 11143 (1999).
\bibitem{kittel}Ch. Kittel, Phys. Rev. \textbf{71}, 270 (1947), Phys. Rev. \textbf{73}, 155
(1948). 
\bibitem{walker}L. R. Walker, Phys. Rev. \textbf{105}, 390 (1957), J. Appl. Phys. \textbf{29},
318 (1958). 
\bibitem{candela 1}D. Candela, N. Masuhara, D. S. Sherill, and D. O. Edwards, J. Low Temp. Phys.
\textbf{63}, 369 (1986). 
\bibitem{candela 2}D. Candela, D. R. McAllaster, and L-J. Wei, Phys. Rev. B \textbf{44}, 7510 (1991). 
\bibitem{Roni}A. Roni and G. Vermeulen, Physica B\textbf{280}, 87 (2000); A. Roni, PhD thesis,
Universit\'{e} Joseph Fourier-Grenoble I, 1999, in French (unpublished, available
in pdf format from \url{www-crtbt.polycnrs-gre.fr/he3pol/dilution/spinwaves.html}). 
\bibitem{table_explain}For susceptibility of the saturated dilute phase we used table II of A. Rodriguez
and G. Vermeulen, J. Low Temp. Phys. \textbf{108}, 103 (1997) based on the susceptibility
data of A. I. Ahonen, M. A. Paalanen, R. C. Richardson, and Y. Takano, J. Low
Temp. Phys. \textbf{24}, 733 (1976) and molar volume and concentration data
of G. E. Watson, J. D. Reppy, and R. C Richardson, Phys. Rev. \textbf{188},
384 (1968). For the susceptibility of pure \( ^{3} \)He, we used a fit to H.
Ramm, P. Pedroni, J. R. Thompson, and H. Meyer, J. Low Temp. Phys. \textbf{2},
539 (1970) and the molar volume given by Greywall, Phys. Rev. B\textbf{33},
7520 (1986). The zero-temperature susceptibility of the very dilute mixtures
is calculated from \( m^{*} \) and \( T_{\mathrm{F}} \) written in Ref.~\cite{candela 2}.
A susceptibility at a finite temperature then was read from the temperature
dependence in Fig.~4 of the same reference. And molar volume data were taken
from Watson et al. Spin diffusion data are taken from the respective sources,
for pure \( ^{3} \)He at 8 bar in \cite{Roni} we used the spin diffusion data
for 6.3 bar from \cite{candela 1}. 
\bibitem{Leggett}A. J. Leggett, J. Phys. C \textbf{3}, 448 (1970). Our notation of \( \mathbf{M} \)
as magnetization --- the local average macroscopic magnetic moment density ---
differs from Leggett's \( \mathbf{M} \) as the local spin density by the factor
\( \gamma \hbar  \). The same holds for the current. 
\bibitem{20}In Leggett's paper the external magnetic field \( \mathbf{H}^{e} \) entered
the equations (\ref{laggett-1}), (\ref{laggett-2}) instead of \( \mathbf{B} \).
Strictly speaking, it should be the exact microscopic field \( \mathbf{h} \),
but the particles motion with ranges greater than interatomic distances leads
to its averaging, giving magnetic flux \( \mathbf{B} \). See discussion at
the end of \S62 of L. D. Landau, E. M. Lifshitz and L. P. Pitaevskii, \emph{Course
of Theoretical Physics}, 2nd edition, Vol. IX, Pergamon Press, 1980. 
\bibitem{Jackson}J. D. Jackson, \emph{Classical Electrodynamics}, 3rd edition, Wiley, NY, 1999. 
\bibitem{LL}L. D. Landau, E. M. Lifshitz and L. P. Pitaevskii, \emph{Course of Theoretical
Physics}, 2nd edition, Vol. VIII, Chapter IX, Pergamon Press, 1984. 
\bibitem{morse}P. M. Morse and H. Feshbach, \emph{Methods of Theoretical Physics,} Vol. I,
McGraw-Hill, NY, 1953. 
\bibitem{GR}I. S. Gradshteyn and I. M. Ryzhik, \emph{Tables of Integrals, Series and Products,}
5th edition, Academic Press, 1994\emph{.}
\bibitem{abramovitz}M. Abramovitz and I. A. Stegun, ed., \emph{Handbook of Mathematical Functions},
9th edition, Dover, NY, 1972. 
\bibitem{merzbacher}E. Merzbacher, \emph{Quantum Mechanics}, 2nd edition, Wiley, NY, 1970. 
\bibitem{patton}The original results of deWames and Wolfram were published rather fragmentarily
in several papers. A good review of the work and all the references may be found
in a review article by C. E. Patton, Phys. Rep. \textbf{103}, 251 (1984). 
\bibitem{Deville}G. Deville, M. Bernier, and J. M. Delrieux, Phys. Rev. B \textbf{19}, 5666 (1979). 
\bibitem{FominVermeulen}I. A. Fomin and G. A. Vermeulen, J. Low Temp. Phys. \textbf{106}, 133 (1997). 
\end{thebibliography}
\end{document}